\documentclass[superscriptaddress,amsmath,amssymb, aps, prx,longbibliography,twocolumn, footinbib]{revtex4-2}

\usepackage{upgreek}
\usepackage{color}
\usepackage{graphicx}
\usepackage{nccmath}
\usepackage{bm}
\usepackage{hyperref}
\hypersetup{colorlinks,breaklinks,
            urlcolor=[rgb]{0,0,0.36},
            linkcolor=[rgb]{0,0,0.36},
            citecolor=[rgb]{0,0,0.36}}
\usepackage[mathlines]{lineno}
\usepackage{dcolumn}

\newcommand{\tr}{\,\mathrm{tr}}
\usepackage{braket}

\newcounter{SN}

\makeatletter
\renewcommand\footnotesize{%
   \@setfontsize\footnotesize\@ixpt{8}%
   \abovedisplayskip 8\p@ \@plus2\p@ \@minus4\p@
   \abovedisplayshortskip \z@ \@plus\p@
   \belowdisplayshortskip 4\p@ \@plus2\p@ \@minus2\p@
   \def\@listi{\leftmargin\leftmargini
               \topsep 4\p@ \@plus2\p@ \@minus2\p@
               \parsep 2\p@ \@plus\p@ \@minus\p@
               \itemsep \parsep}%
   \belowdisplayskip \abovedisplayskip
}
\makeatother

\newcommand{\Harvard}{Department of Physics, Harvard University, Cambridge, Massachusetts, 02138, USA.}

\begin{document}

\title{Emergence of a sharp quantum collective mode in a one-dimensional Fermi polaron}

\author{Pavel~E.~Dolgirev}
\thanks{These authors contributed equally to this work}
\email{p\_dolgirev@g.harvard.edu}
\affiliation{\Harvard}
\author{Yi-Fan~Qu}
\thanks{These authors contributed equally to this work.}
\affiliation{CAS Key Laboratory of Theoretical Physics, Institute of Theoretical Physics, Chinese Academy of Sciences, Beijing 100190, China}
\affiliation{School of Physical Sciences, University of Chinese Academy of Sciences, Beijing 100049, China}

\author{Mikhail~B.~Zvonarev}
\affiliation{Universit{\'e} Paris-Saclay, CNRS, LPTMS, 91405, Orsay, France}
\affiliation{St. Petersburg Department of V.A. Steklov Mathematical Institute of Russian Academy of Sciences, Fontanka 27, St. Petersburg, 191023, Russia}
\affiliation{Russian Quantum Center, Skolkovo, Moscow 143025, Russia}

\author{Tao~Shi}
\email{tshi@itp.ac.cn}
\affiliation{CAS Key Laboratory of Theoretical Physics, Institute of Theoretical Physics, Chinese Academy of Sciences, Beijing 100190, China}
\affiliation{CAS Center for Excellence in Topological Quantum Computation, University of Chinese Academy of Sciences, Beijing 100049, China}

\author{Eugene~Demler}
\affiliation{\Harvard}

\date{\today}

\begin{abstract}
The Fermi-polaron problem of a mobile impurity interacting with fermionic medium emerges in various contexts, ranging from the foundations of Landau's Fermi-liquid theory to electron-exciton interaction in semiconductors, to unusual properties of high-temperature superconductors. While classically the medium provides only a dissipative environment to the impurity, the quantum picture of polaronic dressing is more intricate and arises from the interplay of few- and many-body aspects of the problem. The conventional expectation for the dynamics of Fermi polarons is that it is dissipative in character, and any excess energy is rapidly emitted away from the impurity as particle-hole excitations. Here we report a strikingly different type of polaron dynamics in a one-dimensional system of the impurity interacting repulsively with the fermions. When the total momentum of the system equals the Fermi momentum, there emerges a sharp collective mode corresponding to long-lived oscillations of the polaronic cloud surrounding the impurity. This mode can be observed experimentally with ultracold atoms using Ramsey interferometry and radio-frequency spectroscopy.
\end{abstract}

\maketitle

\section{Introduction}

The problem of a polaron---a mobile particle interacting with a host medium---has a long history dating back to Landau's seminal work on an electron inducing local distortion of a crystal lattice~\cite{landau1933electron}. Polarons are ubiquitous in many-body systems, especially in solid-state~\cite{alexandrov2010advances,devreese_polaron_09} and atomic physics \cite{bloch2012quantum,massignan2014polarons,schmidt2018universal}, and provide one of the key paradigms of modern quantum theory. Recent experimental progress in cold atoms and ion-based quantum simulators brings new motivation for studying polaronic phenomena ~\cite{nascimbene2009collective,schirotzek2009observation,palzer_impurity_transport_09,catani_impurity_dynamics_11,kohstall2012metastability,stojanovic2012quantum,fukuhara_spin_impurity_2013,hu2016bose,jorgensen2016observation,cetina2016ultrafast,meinert2017bloch,yan2019boiling,yan2020bose}, since these platforms offer a high-degree of isolation, tunability of the interaction strength and dispersion, and control of dimensionality~\cite{bloch2008many}.
These setups are particularly well suited for accurate studies of far-from-equilibrium dynamics~\cite{meinert2017bloch,catani_impurity_dynamics_11} since system parameters can be modified much faster than intrinsic timescales of the many-body Hamiltonians. On the theoretical side, recent progress in understanding polarons has come from using such powerful techniques as variational ansatzes~\cite{chevy2006universal,parish2013highly,peotta_mobile_impurity_TDMRG_2013,massel_mobile_impurity_TDMRG_2013,knap2014quantum,li2014variational,shchadilova2016quantum,kain2017hartree,shi2018variational,mistakidis2019repulsive}, renormalization-group calculations~\cite{grusdt2015renormalization,grusdt2018strong}, Monte-Carlo simulations~\cite{prokof2008fermi,ardila2015impurity,grusdt2017bose}, diagrammatic technique~\cite{rosch1995heavy,schmidt2012fermi,rath2013field,burovski_impurity_momentum_2014,gamayun_kinetic_impurity_TG_14,gamayun_quantum_boltzmann_14}, exact Bethe ansatz (BA) calculations for integrable models~\cite{mcguire1965interacting,mcguire1966interacting,yang_fermions_spinful_67,Gamayun_correlation_2016,gamayun_impact_18,gamayun2019zero}, and approaches based on non-linear Luttinger liquids~\cite{imambekov_review_12}. Analysis of equilibrium and dynamical properties of polarons has played an important role in developing new ideas and concepts, and in testing theoretical methods and approaches.

One of the surprising recent discoveries in the far-from-equilibrium dynamics of Fermi polarons has been the prediction of the effect called quantum flutter~\cite{mathy2012quantum,knap2014quantum}: When a repulsive mobile impurity with large momentum is injected into a one-dimensional Fermi gas, it undergoes long-lived oscillations of its velocity. This should be contrasted to the classical situation in which the impurity gradually slows down while transferring its momentum to the host atoms. It has also been found that the quantum flutter frequency does not depend on the initial conditions. The robustness of these oscillations naturally suggests that they represent an incarnation of a fundamental yet unknown property of the polaronic system at equilibrium.

In the present work, we investigate collective modes---elementary excitations describing small deviations from an equilibrium state---in the system of a mobile impurity interacting repulsively with a one-dimensional Fermi gas. Remarkably, we find the density of states of these modes displays a sharp peak when the total momentum of the system equals the Fermi momentum. 
This peak signals the emergence of a distinct collective excitation with a frequency $\omega_{k_F}$, representing a ``breathing mode'' of a polaronic cloud surrounding the impurity. The frequency $\omega_{k_F}$ matches the magnon-plasmon energy-difference at the Fermi momentum, as has been checked for an arbitrary range of model parameters. As we demonstrate below, modern cold-atom techniques, including Ramsey interferometry and radio-frequency (rf) spectroscopy, can be used to detect this mode. Specifically, we find that the impurity absorption spectra at the Fermi momentum exhibit a double-peak structure, with the second peak corresponding to the frequency $\omega_{k_F}$. Our study provides a natural interpretation of such a complex far-from-equilibrium phenomenon as recently discovered quantum flutter in terms of basic equilibrium properties. In particular, we argue that flutter oscillations, as well as their robustness, represent nothing but the signatures of the collective mode $\omega_{k_F}$.

\section{Theoretical framework}
\label{sec:formalism}

A Fermi-polaron model represents a non-trivial many-body problem with the Hamiltonian consisting of three parts: $\hat{H} = \hat{H}_{f} + \hat{H}_{\rm imp} + \hat{H}_{\rm int}$, where $\hat{H}_{f} =  \sum_{k} \frac{k^2}{2m} \hat{c}_{k}^\dagger\hat{c}_{k}$ is the fermionic kinetic energy, $\hat{H}_{\rm imp} = \sum_{k} \frac{k^2}{2M}  \hat{d}_{ k}^\dagger\hat{d}_{k}$ is the kinetic energy of the impurity, and $\hat{H}_{\rm int}=\frac{g}{L} \sum_{k, k', q} \hat{d}^{\dagger}_{k + q}\hat{d}_{k}\hat{c}^{\dagger}_{ k' - q} \hat{c}_{ k'}$ describes contact interaction between the two species of particles. The Planck constant is set to $\hbar = 1$ throughout the paper. Operator $\hat{d}^\dagger_k$ ($\hat{d}_k$) creates (annihilates) the impurity with momentum $k$; operators $\hat{c}^\dagger_k$ and $\hat{c}_k$ represent the host gas. Throughout this work, we assume periodic boundary conditions with the system size $L$, so that $k = \frac{2\pi}{L}n$ with $n$ being integer. The total number of host-gas particles $N$ is fixed via the chemical potential $\mu = \frac{k_F^2}{2m}$, and $k_F =\frac{\pi N}{L}$ is the Fermi momentum. In our calculations, we set $k_F=\frac{\pi}{2}$, which implicitly defines the unit of length. The case of a single impurity restricts the Hilbert space to states with $\sum_{k} \hat{d}_{k}^\dagger\hat{d}_{k} = 1$. The dimensionless interaction strength between the impurity and medium is $\gamma = \frac{ \pi m g}{k_F} $. We use the following convention for the Fourier transform: $\hat{c}_x = \frac{1}{\sqrt{L}} \sum_k e^{ikx}\hat{c}_k$. We choose a sufficiently large UV momentum cutoff $\Lambda\gg k_F$ in our numerical simulations.

A challenge one encounters when solving a many-body problem is that the Hilbert space grows exponentially with the system size, limiting direct numerical simulations to relatively small systems. One approach to overcoming this difficulty is to employ a variational method, where a limited number of parameters is used to parameterize a class of many-body states. In this approach, the complexity of computations typically grows polynomially with the system size, allowing for efficient numerical analysis. However, one needs to ensure that the variational wave function contains the right class of quantum states that can reliably capture the many-body correlations. More specifically, a variational family of states is required to satisfy the following criteria: (i) it contains a manageable number of variational parameters, (ii) it accurately predicts ground-state properties, (iii) it captures real-time dynamics including the spectrum of collective modes, and (iv) it can be used to compute observables relevant for experiments.

We employ recent developments of approaches based on non-Gaussian states (NGS) to realize this program~\cite{shi2018variational,hackl2020geometry}. In this work, we deal with zero-temperature situations, and finite-temperature ensembles can be studied using the formalism developed in Ref.~\cite{shi2019variational}. For the Fermi-polaron problem, one first performs a unitary transformation to the impurity reference frame~\cite{lee1953motion,kain2017hartree}, $\hat{{\cal S}} = \exp(- i \hat{x}_{\rm imp} \hat{ P}_f )$, where $\hat{P}_f = \sum_{ k}  k \,\hat{c}^\dagger_{ k} \hat{c}_{k}$ is the total fermionic momentum and $\hat{x}_{\rm imp}$ is the impurity position operator, and then invokes the Hartree-Fock approximation. The unitary transformation $\hat{{\cal S}}$ plays a two-fold role: (i) it provides sufficient entanglement between the impurity and the medium so that the Hartree-Fock approximation becomes accurate, and (ii) it takes advantage of the total momentum conservation and decouples the impurity from the rest of the system. In the impurity frame, the transformed Hamiltonian is parametrized by the total momentum $Q$:
\begin{equation}
    \hat{H}_{Q} = \sum_{ k, k'}\hat{c}_{k}^\dagger \left[\frac{k^2}{2m}\delta_{k k'}  + \frac{g}{L}\right] \hat{c}_{k'} + \frac{(Q - \hat{P}_f)^2}{2M}. \label{eqn::LLP_H}
\end{equation}
Note that only the degrees of freedom of the host gas enter Eq.~\eqref{eqn::LLP_H}. The first term is the fermionic kinetic energy, the second term describes scattering off the impurity, and the third term corresponds to its recoil energy.

\subsection{Equations of motion}

The Hartree-Fock approximation can be conveniently cast into a Gaussian wave function~\cite{shi2018variational}, which we write as
\begin{equation}
    \Ket{\psi (t)} = {\rm e}^{-i\theta} \exp( i \hat{c}^\dagger \xi \hat{c}) \Ket{\rm FS}, \label{eqn:wf}
\end{equation}
where $\xi = \xi^\dagger$ and $\Ket{\rm FS} \equiv \prod_{|k|\leq k_F} \hat{c}^\dagger_k\Ket{0}$ is the wave function of the filled Fermi sea ($\Ket{0}$ corresponds to the vacuum state). The information about the state~\eqref{eqn:wf} is then encoded in $\theta$ and $U \equiv {\rm e}^{i\xi}$. Let us define the covariance matrix as:
$\Gamma_{k,k'} \equiv \langle \hat{c}^\dagger_k \hat{c}_{k'} \rangle_{\rm GS} = U^* \Gamma_0 U^T,$
where $\Gamma_0$ is the covariance matrix of the filled Fermi sea. We emphasize that even though we choose the state to be Gaussian in the impurity reference frame, it is non-Gaussian in the laboratory frame due to the unitary operator ${\cal \hat{S}}$.

To find the (momentum-dependent) ground state wave function, we employ the imaginary-time dynamics~\cite{shi2018variational}:
\begin{equation}
d_{\tau} \Gamma = 2 \Gamma h \Gamma - \left\{ h, \Gamma \right\},\label{eqn:im_Gamma}
\end{equation}
where 
\begin{align}
   E[\Gamma] & \equiv \langle \hat{H}_{ Q}\rangle = \sum_{k,k'} \Gamma_{kk'}\left[ \Big(\epsilon_k +\frac{k^2}{2M} -\frac{Q\cdot k}{M}\Big)\delta_{kk'} + \frac{g}{L} \right] \notag\\
   & \qquad\qquad - \sum_{k,k'}\frac{k \cdot k'}{2M} |\Gamma_{k k'}|^2 +\frac{P_f^2}{2M} + \frac{Q^2}{2M}, \\
    h_{k k'} & \equiv  \frac{\delta E[\Gamma]}{\delta \Gamma_{k'k}} = \Big(\epsilon_k + \frac{k^2}{2M} + \frac{k \cdot ( P_f -  Q)}{M}   \Big) \delta_{k k'} \notag\\
    & \qquad \qquad \qquad \qquad\qquad
    +\frac{g}{L}-\frac{k\cdot  k'}{M}\Gamma_{ k k'}.\label{eqn::h}
\end{align}
Here $\epsilon_k = \frac{k^2}{2m} - \mu$. Note that initially pure states, with $\Gamma^2 = \Gamma$, will remain pure under the imaginary-time evolution: $d_{\tau} (\Gamma^2 - \Gamma) = 0$. For these states, the total number of fermions $N = \sum_k {\Gamma}_{kk}$ is conserved.

The equations of motion for the real-time dynamics are obtained from the Dirac's variational principle:
\begin{equation}
    \partial_t \theta = E[\Gamma] - \tr h\, \Gamma,\quad i\partial_t U  = h^* U. \label{eqn::dt_U}
\end{equation}
From Eqs.~\eqref{eqn::dt_U} one can derive an equation of motion solely on the covariance matrix~\cite{shi2018variational}:
\begin{equation}
    d_t \Gamma = i \left[ h,\Gamma \right]. \label{eqn::real_dyn_Gamma}
\end{equation}
During the real-time dynamics, the state remains pure, and the total number of fermions is also conserved.

\subsection{Analysis of collective modes}
\label{subsec:CM}

Collective excitations represent low-energy small-amplitude fluctuations on top of an equilibrium state, in our case on top of a ground state, with the covariance matrix $\Gamma_Q$, previously computed via the imaginary-time dynamics. To obtain their spectrum within the NGS approach, we write $\Gamma = \Gamma_Q + \delta \Gamma$, and, assuming that $\delta \Gamma$ is small, we linearize the real-time equation of motion~\eqref{eqn::real_dyn_Gamma}:
\begin{align}
    d_t  \delta \Gamma &= i\left[ h_Q,\delta \Gamma \right] + \frac{i}{M}\left[\tr({P}\delta \Gamma) {P} - {P} \delta \Gamma {P},\Gamma_Q \right], \label{eqn::dt_delta_Gamma}
\end{align}
where $h_Q = h[\Gamma_Q]$ and ${P}_{k,k'} \equiv k\delta_{k,k'}$. Importantly, these fluctuations $\delta \Gamma$ are constrained to satisfy: i) hermiticity $\delta \Gamma = \delta \Gamma^\dagger$, ii) particle number conservation $\tr \, \delta \Gamma = 0$, and iii) purity $\{ \Gamma_Q, \delta \Gamma \} = \delta \Gamma$. 
To see the implications of these conditions, we switch to the basis where the matrix $\Gamma_Q$ is diagonal:
\begin{align}
    \Gamma_Q = \begin{pmatrix} 0 & 0\\
0 & I_{N\times N}
\end{pmatrix} \Longrightarrow \delta \Gamma = \begin{pmatrix}
0 & K\\
K^\dagger & 0
\end{pmatrix}, \label{eqn:K_def}
\end{align}
where $K$ is an arbitrary $(N_{\rm sp}-N)\times N$ matrix that fully parametrizes all possible physical fluctuations $\delta \Gamma$. $N_{\rm sp}$ is the total number of single-particle modes in the fermionic system, and it is determined by the system size $L$ and by the UV-cutoff $\Lambda$. We, therefore, conclude that the above constraints reduce the total number of degrees of freedom in $\delta \Gamma$ from $2N_{\rm sp}^2$ to only $2N(N_{\rm sp}-N)$. Plugging in Eq.~\eqref{eqn:K_def} into Eq.~\eqref{eqn::dt_delta_Gamma}, one obtains:
\begin{align}
    d_t K & = i (h_{11}K - K h_{22}) + \frac{i}{M} P_{12} \tr( P_{12} K^\dagger + P_{21} K ) \notag \\
    &\qquad - \frac{i}{M} (P_{11}KP_{22} + P_{12}K^\dagger P_{12}), \label{eqn::dt_K}
\end{align}
where we used that in the chosen basis $h_Q = \begin{pmatrix}
h_{11} & 0\\
0 & h_{22}
\end{pmatrix}$, since $h_Q$ and $\Gamma_Q$ commute (in principle, one can always choose a basis where both $h_Q$ and $\Gamma_Q$ are diagonal). The eigenenergies $\omega_i^Q$ of Eq.~\eqref{eqn::dt_K} constitute the spectrum of collective excitations (here $Q$ is the total momentum of the system, and $i$ labels excitations). From $\omega_i^Q$ and corresponding eigenvectors, one can compute standard linear response functions such as the density response function~\cite{kamenev2011field}. For a bosonic system, similar fluctuation analysis has been proven to be equivalent to the generalized random phase approximation~\cite{guaita2019gaussian} and successfully applied to reproduce the Goldstone zero-mode naturally without imposing the Hugenholtz-Pines condition.

Using the outlined theoretical framework, we obtained our main results, which we turn to discuss in the next section.

\begin{figure*}[t!]
\centering
\includegraphics[width=1\linewidth]{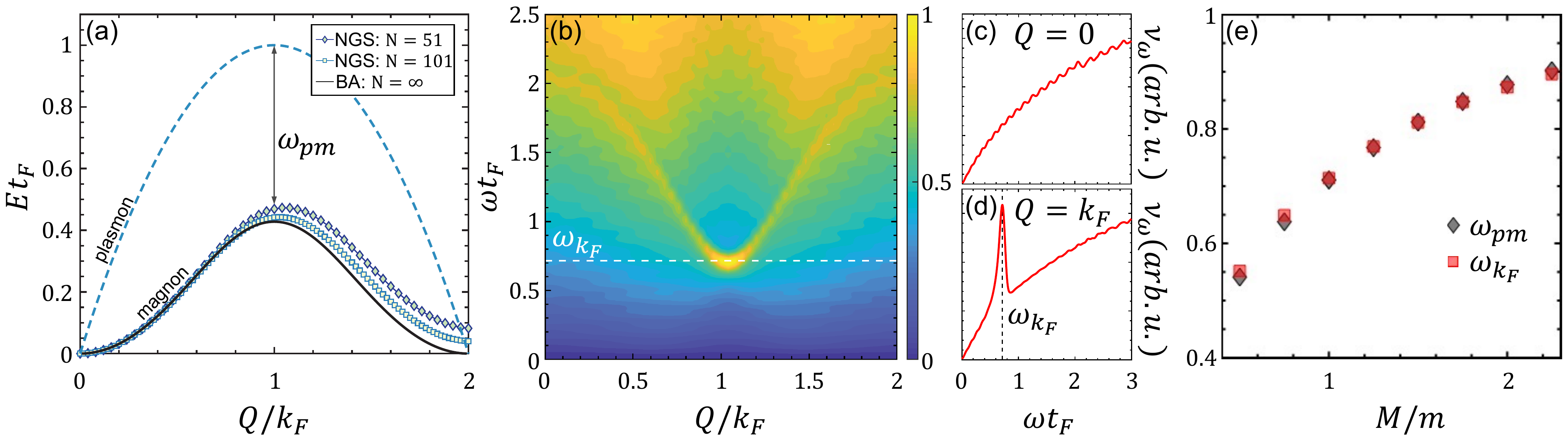} 
\caption{(a) Polaron energy-momentum relation---the magnon branch---for the case of equal masses $M = m$ and the plasmon branch of the host Fermi sea. Note the plasmon-magnon excitation energy $\omega_{\rm pm}$ at $k_F$. Our variational NGS approach remarkably reproduces the exact BA result (solid line), adopted from Refs.~\cite{mcguire1965interacting,gamayun2019zero} (here we used $\Lambda = 10 k_F$). (b) Density of states (DOS) of collective excitations as a function of frequency $\omega$ and total momentum $Q$. Here $t_F = 1/E_F$ is the Fermi time. Note that the spectral signal becomes particularly pronounced at $Q = k_F$. (c) and (d) Cuts of the DOS at $Q=0$ and $Q=k_F$, respectively. Dashed line in (d) (shown also in (b)) indicates the emergence of a sharp mode. (e) The collective mode $\omega_{k_F}$, as a function of the mass ratio $M/m$, matches the plasmon-magnon mode $\omega_{\rm pm}$ from (a). Parameters used: $\gamma = 5$, $N = 51$, and $\Lambda = 5k_F$.}
\label{fig::Main}
\end{figure*}

\section{Main Results and Discussion}
\label{sec:results}

\subsection{Emergent collective mode and its origin}

An important feature of the polaron system is momentum conservation, which after the Lee-Low-Pines transformation and via the imaginary-time dynamics, allows us to compute not only the ground-state energy but rather the entire polaron energy-momentum dispersion. An example of such a calculation for the integrable case of equal masses $M = m$ is shown in Fig.~\ref{fig::Main} (a), where we compare momentum-dependent ground-state energies to the exact Bethe ansatz (BA) results~\cite{mcguire1965interacting,gamayun2019zero}. The agreement is excellent, and it becomes even better for a larger total number of particles $N$ and/or larger momentum cutoff $\Lambda$. Note that in the thermodynamic limit $N\to\infty$, this energy-momentum relation is $2k_F$-periodic, since at $Q = 2k_F$, one can always excite a zero-energy particle-hole pair across the Fermi surface. At finite $N$, this is no longer true, and to excite such a pair costs energy proportional to $1/N$, explaining the discrepancy in Fig.~\ref{fig::Main} (a) at large momenta $Q\simeq 2k_F$. In Appendix~\ref{Appendix_BA_static}, we demonstrate that our approach also reproduces exact many-body correlation functions. Below we also investigate a generic situation of not equal masses, where no known exact solutions are available.

Using the approach outlined in Sec.~\ref{subsec:CM}, we turn to investigate the spectrum of collective modes on top of momentum-dependent ground states $\Gamma_Q$. Figure~\ref{fig::Main}~(b) shows the density of states (DOS) of these excitations, $\nu_\omega = \sum_i \delta(\omega - \omega_i^Q)$. 
Most spectacularly, we discover a sharp peak at $Q = k_F$ [Fig.~\ref{fig::Main}~(c)], which signals the onset of a new distinct collective mode. Our primary goal below is to elucidate its physical origin and investigate the feasibility of experimental verification with ultracold atoms.

\begin{figure}[t!]
\centering
\includegraphics[width=1\linewidth]{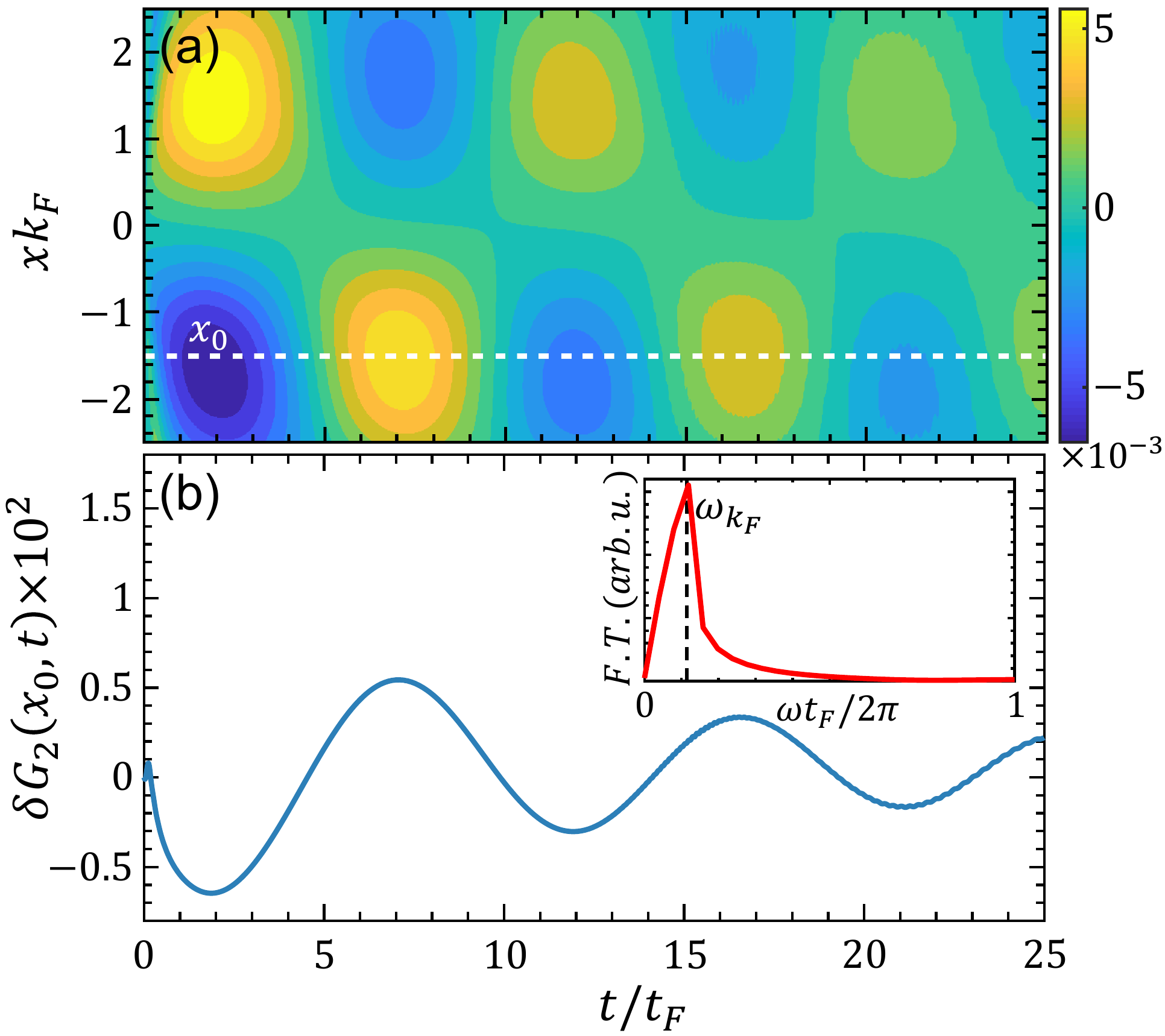} 
\caption{Linear-response dynamics in the impurity frame after a soft quench of the coupling strength $\gamma$: from $\gamma = 6$ to $\gamma = 5$. The wave function at $t=0$ corresponds to the ground state at $Q = k_F$ and $\gamma = 6$. (a) Evolution of $\delta G_2(x,t) = G_2(x,t)-G_2(x,0)$ showing oscillatory behavior of the fermionic density surrounding the impurity. (b) Dynamics of $\delta G_2(x,t)$ at $x = x_0$ (dashed line in (a)). Notably, the Fourier transform of this signal (inset) matches the frequency $\omega_{k_F}$ extracted from Fig.~\ref{fig::Main}~(b).}
\label{fig::G2_xt}
\end{figure}

We turn to discuss the physical mechanism behind the emergence of this mode $\omega_{k_F}$. We first identify which states in the many-body spectrum determine the frequency $\omega_{k_F}$. Let us define plasmon as the lowest energy excitation of the Fermi gas in the absence of impurity. Its dispersion, $E_p(Q)$, has a familiar inverse-parabolic shape, shown in Fig.~\ref{fig::Main}~(a) with dashed line. Magnon is the lowest energy excitation of the entire interacting system. The magnon dispersion $E_m(Q)$---the polaron energy-momentum relation---is illustrated in Fig.~\ref{fig::Main}~(a) with a solid line. In the presence of the impurity, the plasmon still exists, but no longer represents the lowest-energy excitation. Note that both magnon and plasmon group velocities evaluated at $Q=k_F$---at the same wave vector where the mode $\omega_{k_F}$ emerges---are zero, suggesting that these two states can form a correlated long-lived excitation, with frequency $\omega_\mathrm{pm}(k_F) = E_p(k_F)-E_m(k_F)$. Interestingly, our variational calculations show that
\begin{equation}
    \omega_{k_F} = \omega_\mathrm{pm}(k_F)
\end{equation}
for any impurity-gas mass ratios $M/m$ and coupling strengths $\gamma$, as illustrated in Fig.~\ref{fig::Main}~(e).

Now we demonstrate that the collective excitation $\omega_{k_F}$ represents oscillations of the polaronic cloud surrounding the impurity. To that end, we take the initial many-body wave function \textbf{$|\psi_\mathrm{lab}(0)\rangle = |\mathrm{GS}_Q\rangle$} to be the ground state of the interacting Fermi polaron model with the total momentum $Q = k_F$, and then suddenly change the interaction strength. In response to such a quench, we find that the fermionic density in the vicinity of the impurity
\begin{align*}
G_2(x,t)= \frac{L}{N} \int\limits_0^L dy  \Bra{\psi_{\rm lab}(t)} \hat{d}^\dagger_y\hat{d}_y \hat{c}^\dagger_{x+y}\hat{c}_{x + y}\Ket{\psi_{\rm lab}(t)} \label{eqn:G2_xt}
\end{align*}
demonstrates damped oscillatory behavior, illustrated in Fig.~\ref{fig::G2_xt}, with the frequency $\omega_{k_F}$.  These real-time dynamical correlations are, in principle, accessible with ultracold-atom setups. One can see, however, that the amplitude of the signal shown in Fig.~\ref{fig::G2_xt} is rather small because the system is close to the linear-response regime.  We find that the amplitude of oscillations remains small even for stronger quenches. To overcome this issue, below we suggest a complementary experimental verification of our findings by computing observables accessible with rf spectroscopy and Ramsey-type interferometry.

\begin{figure*}[htb!]
\centering
\includegraphics[scale=0.44]{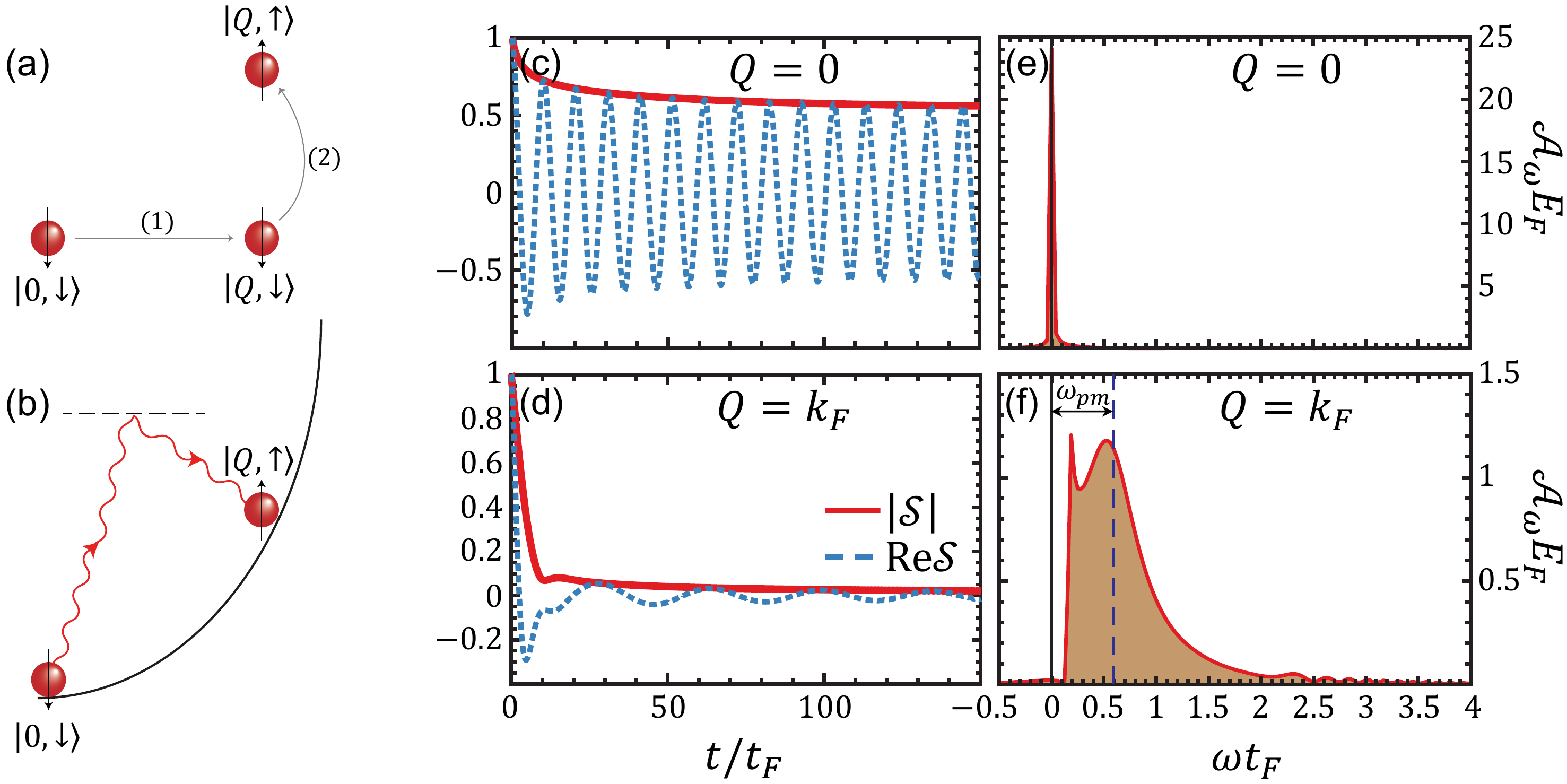} 
\caption{(a) and (b) Possible cold-atom setups. The initial wave function corresponds to $\Ket{\rm FS}\otimes\Ket{0,\downarrow}$, where the hyperfine state $\Ket{\downarrow}$ does not interact with the fermionic medium. (a) The impurity is first accelerated such that it acquires momentum $Q$; subsequent rf-pulse drives it into the hyperfine state $\Ket{\uparrow}$ strongly interacting with the host gas. (b) Alternatively, the two states $\Ket{0,\downarrow}$ and $\Ket{Q,\uparrow}$ can be directly coupled by a two-photon Raman process. (c) to (f) The dynamical overlap function ${\cal S}(t)$ and the impurity absorption spectra ${\cal A}_\omega$ for $Q = 0$ (panels (c) and (e)) and for $Q = k_F$ (panels (d) and (f)). We shifted frequencies in (e) and (f) such that the zero value in both panels represents the corresponding ground-state oscillations. For $Q = k_F$, the Ramsey contrast $|{\cal S}(t)|$ demonstrates switching from initial rapid decay for times $t\lesssim 10 t_F$ to the lasting regime of slow dynamics. This behavior reflects in ${\cal A}_\omega$ as it acquires a double-peak structure: The frequency of the first peak is close to that of the ground state oscillations, whereas the second peak corresponds to the collective mode $\omega_{k_F}=\omega_{\rm pm}$. Parameters are the same as in Fig.~\ref{fig::Main}, except $N = 251$.}
\label{fig::Combined}
\end{figure*}

\subsection{Cold-atom setups}

Possible experimental setups for investigating the physics of a mobile impurity coupled to a Fermi bath are shown in Fig.~\ref{fig::Combined}~(a) and~(b). We assume that the impurity has two hyperfine states: $\Ket{\downarrow}$ is decoupled from the Fermi sea, whereas $\Ket{\uparrow}$  strongly interacts with the host gas. We start from an initial many-body wave function prepared in the state $\Ket{\rm FS}\otimes\Ket{0,\downarrow}$. $\Ket{Q,\downarrow}$ labels the impurity state with the total momentum $Q$. To reach a given total momentum sector $Q$, we suggest the quenching protocol illustrated in Fig.~\ref{fig::Combined}~(a). The impurity is first accelerated---for example, by application of an external force as in Ref.~\cite{meinert2017bloch}---such that its momentum becomes $Q$. Then an rf-pulse is used to couple the two hyperfine states. Similar to the case of a static impurity discussed in Ref.~\cite{knap2012time}, Ramsey interferometry can probe the dynamical overlap function, which in our case is written as ${\cal S}(t) = \Bra{\rm FS} {\rm e}^{it\hat{H}_Q^{(0)}}{\rm e}^{-it\hat{H}_Q} \Ket{\rm FS}$, where $\hat{H}_Q^{(0)}$ is given by Eq.~\eqref{eqn::LLP_H} with $g = 0$. The impurity absorption spectra is obtained as ${\cal A}_\omega = \frac{1}{\pi} \text{Re} \int_0^{\infty}dt\,{\rm e}^{i\omega t} {\cal S}(t)$. Figure~\ref{fig::Combined}~(b) shows an alternative experimental setup~\cite{ness2020observation}, where one employs the two-photon Bragg spectroscopy~\cite{stenger1999bragg,stamper1999excitation,ozeri2005colloquium}. In this latter situation, the dynamical overlap function is modified by a non-essential phase factor.

The overlap function ${\cal S}(t)$ is computed analytically:
\begin{equation}
    {\cal S}(t) = {\rm e}^{-i(\theta(t)-E[\Gamma_0]t)} \det (1 - (1-U(t))\Gamma_0^T).
\end{equation}
This expression is a generalization of the approach used for the case of static impurity~\cite{knap2012time}. A numerical simulation of Eqs.~\eqref{eqn::dt_U} indicates that ${\cal S}(t)$ exhibits long-time revivals (roughly at $t \simeq L/k_F$) associated with the finite system size $L$. Below we, therefore, choose a sufficiently large system such that these revivals do not appear up to the largest simulation times.

For the case $Q=0$, shown in Fig.~\ref{fig::Combined}~(c) and~(e), the Ramsey contrast $\lvert{\cal S}(t)\rvert$ demonstrates a slow monotonic decay; at long times it saturates around $R_0 = \lvert \braket{\rm GS_0| \rm FS}\rvert^2$, which equals $R_0 \simeq 0.6$ for the parameters used in Fig.~\ref{fig::Combined}. Note, however, that in the thermodynamic limit, $L\to \infty$, this quasiparticle residue $R_0$ should vanish -- an analog of the Anderson orthogonality catastrophe for the case of a static impurity. This latter statement we explicitly verify numerically in Appendix~\ref{Appendix_Residue}, where we show that $R_0$ decays with the system size $L$ as a power-law. We note in passing that this decay is much slower than in the case $M=\infty$. In the frequency domain, ${\cal A}_\omega$ displays a sharp peak at the polaron binding energy $E_0$, as it should be. 

Figure~\ref{fig::Combined}~(d) and~(f) shows the results for $Q=k_F$. We find that ${\cal S}(t)$ demonstrates a qualitative change in its dynamics roughly at $t\simeq 10t_F$: The initial quick decay of $|{\cal S}(t)|$, associated with the fact that the initial wave function represents a far-from-equilibrium state for $Q=k_F$, turns into a much slower power-law decay at longer times. This behavior resembles the non-Markovian spontaneous emission of a two-level atom coupled to a non-flat photon bath~\cite{cohen1998atom}, in which case the initial fast decay is associated with the large DOS of the collective mode, cf. Fig.~\ref{fig::Main}~(b). For the impurity absorption spectra ${\cal A}_\omega$, we find it acquires a double-peak structure. 
Importantly, the second broad peak corresponds to the collective mode $\omega_{k_F}$ -- the dashed line in Fig.~\ref{fig::Combined} (f) denotes the discussed plasmon-magnon mode $\omega_{\rm pm}(k_F)$. The position of the first peak is close to the frequency of the ground state at $Q=k_F$. There are a few reasons for the small mismatch between them. First, we find that the overlap $R_{k_F}= \lvert \braket{{\rm GS}_{k_F}| \rm FS}\rvert^2$ is suppressed: it equals $4\times 10^{-3}$ for the parameters used in Fig.~\ref{fig::Combined}. Therefore, the first maximum in ${\cal A}_\omega$ is shifted towards higher frequencies, where the overlap of the initial wave function and an excited state is more pronounced. Second, such a small value of $R_{k_F}$ further indicates that the intrinsic dynamics is far-from-equilibrium. However, in an out-of-equilibrium setting, our method is not expected to be exact. Indeed, an explicit comparison in Appendix~\ref{Appendix_dyn} of dynamics in the NGS approach to that of the BA indicates that our method displays a similar small discrepancy with the exact result. Qualitatively, the method provides correct predictions even for non-equilibrium problems. Finally, the small mismatch could potentially be reduced by increasing the frequency resolution, which requires a simulation of an even larger system and for longer times.

\subsection{Quantum flutter and its robustness}

Equipped with our main results outlined above, we turn to discuss the phenomenon of quantum flutter. While the quantum flutter embodies a strongly out-of-equilibrium character, its phenomenology, surprisingly, shares a lot in common with the discovered here equilibrium collective mode $\omega_{k_F}$. Indeed, essentially arbitrary quenching of the polaronic system~\cite{mathy2012quantum,knap2014quantum} results in the development at long times of long-lived oscillations of the polaronic cloud surrounding the impurity. For the integrable situations, it was shown that the frequency of those oscillations matches the plasmon-magnon energy difference at $k_F$. The robustness of quantum flutter oscillations to quenching conditions, together with the results of our work, suggests the following interpretation. The initial (strong) perturbation of the polaronic system results in exciting all kinds of possible excitations. Only the ones with small relative group velocities contribute to the impurity dynamics at long times, while the role of the rest of them is to carry energy and momentum away from the impurity. Although the system is out-of-equilibrium, near the impurity, and at long times, the dynamics should remind the equilibrium situation, such as in Fig.~\ref{fig::G2_xt}. This is i) because  both the plasmon and magnon group velocities are zero at $k_F$ and ii) due to the enhanced density of states of the associated collective mode $\omega_{k_F}$, cf. Fig.~\ref{fig::Main}~(b). This physical picture explains long-time flutter oscillations and their robustness.

We turn to discuss new insights that our work brings to the physics of quantum flutter. The plasmon and magnon at $k_F$ form a long-lived collective mode $\omega_{k_F}$, although the plasmon lives in the continuum of excitations of the interacting system. The sharp collective mode is responsible for a robust, long-lived oscillatory dynamics, independent of quenching conditions. More generally, we expect the development of such a collective mode provided there exist two branches of excitations, which are required to have zero group velocities simultaneously (as such, the corresponding density of states are enhanced due to Van Hove singularities). In contrast to existing theoretical approaches, our variational states enable an accurate computation of the magnon dispersion away from the integrable points. We remark that the plasmon-magnon energy difference at $k_F$ shown in Fig.~\ref{fig::Main}~(e) is consistent with the flutter frequency for $M \geq m$ obtained from DMRG simulations~\cite{knap2014quantum}. In the regime of light impurity, $M \lesssim m$, the two approaches start to deviate, which warrants further investigation. A probable reason for this is that it becomes notoriously difficult to extract the flutter frequency reliably in this regime (see Appendix~\ref{appendix:CM}).

\section{Conclusion and Outlook}
\label{sec:discussion}

The Fermi polaron Hamiltonian represents an instance of an interacting many-body system without a small parameter. As such, to establish the validity of any conclusions, it is crucial to verify that our chosen variational states are sufficient to capture both equilibrium and out-of-equilibrium properties. Importantly, it turns out that the case $M=m$ is Bethe ansatz solvable, providing us with the necessary ground for testing our approach. Specifically, in Appendix~\ref{Appendix_BA_static}, we demonstrate that our non-Gaussian states reproduce both the ground-state energies and many-body correlation functions, both for repulsive and attractive interactions. Benchmarking the out-of-equilibrium situation is more challenging. In Appendix~\ref{Appendix_dyn}, we show that the developed here variational approach captures the essential features of the quantum flutter dynamics, although not perfectly. Our work, therefore, establishes the reliability of the non-Gaussian states in one spatial dimension, where fluctuations are expected to be the strongest. Because of efficient numerical implementation, these variational states are promising to solve interacting many-body problems in higher dimensions, where no such powerful numerical methods as DMRG exist, with Monte Carlo-based approaches being the only alternative. They can be used for computing ground-state properties, including many-body correlations and collective modes, as well as out-of-equilibrium real-time dynamics.

Once we established the reliability of the non-Gaussian approach, we turned to investigate collective excitations of the polaronic system at equilibrium. Our main result is that their spectrum contains a sharp peak when the total momentum of the system equals $k_F$, signaling the onset of a new distinct collective mode. By analyzing polaron dispersion for an arbitrary range of the mass ratios $M/m$ and interaction strengths $\gamma$, we concluded that this mode represents nothing but the plasmon-magnon excitation at $k_F$. Our work suggests a close connection between the far-from-equilibrium Fermi-polaron dynamics and equilibrium collective excitations. This relation explains the robustness of the phenomenon of quantum flutter to changes in model parameters and initial conditions. Theoretical predictions made in this paper can be tested with currently available experimental systems of ultracold atoms. Specifically, one can search for the following features that should appear when the momentum of the impurity relative to the host atoms reaches $k_F$: (i) long-lived oscillations of the polaronic cloud, (ii) abrupt change in the time evolution of the Ramsey contrast, and (iii) development of the double-peak structure in the impurity absorption spectra at the Fermi momentum. The frequency of oscillations can be tuned experimentally by varying either the interaction strength $\gamma$ or the mass ratio $M/m$. From a broader perspective, our work is inspired by the recent developments in designing and studying controlled quantum systems using both solid-state and cold-atom platforms. In particular, as modern semiconductor technologies are approaching the quantum domain with the current feature-size of a few nanometers, understanding far-from-equilibrium dynamics of interacting fermionic systems will be crucial for the design and operation of future electronic devices.

\section{Acknowledgements}
\noindent We thank I.~Bloch, M.~Zwierlein, Z.~Yan, I.~Cirac, D.~Wild, F. Grusdt, K. Seetharam, O.~Gamayun, and D.~Sels for fruitful discussions. P.~E.~D. and E.~D. are supported by the Harvard-MIT Center of Ultracold Atoms, ARO grant number W911NF-20-1-0163, and NSF EAGER-QAC-QSA. T.~S. is supported by the NSFC (Grants No. 11974363). The work of M.~B.~Z. is supported by Grant No.~ANR-16-CE91-0009-01 and CNRS grant PICS06738.

\bibliography{polaron}

\begin{thebibliography}{60}%
\makeatletter
\providecommand \@ifxundefined [1]{%
 \@ifx{#1\undefined}
}%
\providecommand \@ifnum [1]{%
 \ifnum #1\expandafter \@firstoftwo
 \else \expandafter \@secondoftwo
 \fi
}%
\providecommand \@ifx [1]{%
 \ifx #1\expandafter \@firstoftwo
 \else \expandafter \@secondoftwo
 \fi
}%
\providecommand \natexlab [1]{#1}%
\providecommand \enquote  [1]{``#1''}%
\providecommand \bibnamefont  [1]{#1}%
\providecommand \bibfnamefont [1]{#1}%
\providecommand \citenamefont [1]{#1}%
\providecommand \href@noop [0]{\@secondoftwo}%
\providecommand \href [0]{\begingroup \@sanitize@url \@href}%
\providecommand \@href[1]{\@@startlink{#1}\@@href}%
\providecommand \@@href[1]{\endgroup#1\@@endlink}%
\providecommand \@sanitize@url [0]{\catcode `\\12\catcode `\$12\catcode
  `\&12\catcode `\#12\catcode `\^12\catcode `\_12\catcode `\%12\relax}%
\providecommand \@@startlink[1]{}%
\providecommand \@@endlink[0]{}%
\providecommand \url  [0]{\begingroup\@sanitize@url \@url }%
\providecommand \@url [1]{\endgroup\@href {#1}{\urlprefix }}%
\providecommand \urlprefix  [0]{URL }%
\providecommand \Eprint [0]{\href }%
\providecommand \doibase [0]{https://doi.org/}%
\providecommand \selectlanguage [0]{\@gobble}%
\providecommand \bibinfo  [0]{\@secondoftwo}%
\providecommand \bibfield  [0]{\@secondoftwo}%
\providecommand \translation [1]{[#1]}%
\providecommand \BibitemOpen [0]{}%
\providecommand \bibitemStop [0]{}%
\providecommand \bibitemNoStop [0]{.\EOS\space}%
\providecommand \EOS [0]{\spacefactor3000\relax}%
\providecommand \BibitemShut  [1]{\csname bibitem#1\endcsname}%
\let\auto@bib@innerbib\@empty
\bibitem [{\citenamefont {Landau}(1933)}]{landau1933electron}%
  \BibitemOpen
  \bibfield  {author} {\bibinfo {author} {\bibfnamefont {L.~D.}\ \bibnamefont
  {Landau}},\ }\bibfield  {title} {\bibinfo {title} {Electron motion in crystal
  lattices},\ }\href@noop {} {\bibfield  {journal} {\bibinfo  {journal} {Phys.
  Z. Sowjet.}\ }\textbf {\bibinfo {volume} {3}},\ \bibinfo {pages} {664}
  (\bibinfo {year} {1933})}\BibitemShut {NoStop}%
\bibitem [{\citenamefont {Alexandrov}\ and\ \citenamefont
  {Devreese}(2010)}]{alexandrov2010advances}%
  \BibitemOpen
  \bibfield  {author} {\bibinfo {author} {\bibfnamefont {A.~S.}\ \bibnamefont
  {Alexandrov}}\ and\ \bibinfo {author} {\bibfnamefont {J.~T.}\ \bibnamefont
  {Devreese}},\ }\href@noop {} {\emph {\bibinfo {title} {Advances in polaron
  physics}}},\ Vol.\ \bibinfo {volume} {159}\ (\bibinfo  {publisher}
  {Springer},\ \bibinfo {year} {2010})\BibitemShut {NoStop}%
\bibitem [{\citenamefont {Devreese}\ and\ \citenamefont
  {Alexandrov}(2009)}]{devreese_polaron_09}%
  \BibitemOpen
  \bibfield  {author} {\bibinfo {author} {\bibfnamefont {J.~T.}\ \bibnamefont
  {Devreese}}\ and\ \bibinfo {author} {\bibfnamefont {A.~S.}\ \bibnamefont
  {Alexandrov}},\ }\bibfield  {title} {\bibinfo {title} {Fr\"ohlich polaron and
  bipolaron: recent developments},\ }\href
  {https://doi.org/10.1088/0034-4885/72/6/066501} {\bibfield  {journal}
  {\bibinfo  {journal} {Rep. Prog. Phys.}\ }\textbf {\bibinfo {volume} {72}},\
  \bibinfo {pages} {066501} (\bibinfo {year} {2009})},\ \Eprint
  {https://arxiv.org/abs/arXiv:0904.3682} {arXiv:0904.3682} \BibitemShut
  {NoStop}%
\bibitem [{\citenamefont {Bloch}\ \emph {et~al.}(2012)\citenamefont {Bloch},
  \citenamefont {Dalibard},\ and\ \citenamefont
  {Nascimbene}}]{bloch2012quantum}%
  \BibitemOpen
  \bibfield  {author} {\bibinfo {author} {\bibfnamefont {I.}~\bibnamefont
  {Bloch}}, \bibinfo {author} {\bibfnamefont {J.}~\bibnamefont {Dalibard}},\
  and\ \bibinfo {author} {\bibfnamefont {S.}~\bibnamefont {Nascimbene}},\
  }\bibfield  {title} {\bibinfo {title} {Quantum simulations with ultracold
  quantum gases},\ }\href {https://doi.org/10.1038/nphys2259} {\bibfield
  {journal} {\bibinfo  {journal} {Nat. Phys.}\ }\textbf {\bibinfo {volume}
  {8}},\ \bibinfo {pages} {267} (\bibinfo {year} {2012})}\BibitemShut {NoStop}%
\bibitem [{\citenamefont {Massignan}\ \emph {et~al.}(2014)\citenamefont
  {Massignan}, \citenamefont {Zaccanti},\ and\ \citenamefont
  {Bruun}}]{massignan2014polarons}%
  \BibitemOpen
  \bibfield  {author} {\bibinfo {author} {\bibfnamefont {P.}~\bibnamefont
  {Massignan}}, \bibinfo {author} {\bibfnamefont {M.}~\bibnamefont
  {Zaccanti}},\ and\ \bibinfo {author} {\bibfnamefont {G.~M.}\ \bibnamefont
  {Bruun}},\ }\bibfield  {title} {\bibinfo {title} {Polarons, dressed molecules
  and itinerant ferromagnetism in ultracold fermi gases},\ }\href
  {https://doi.org/10.1088/0034-4885/77/3/034401} {\bibfield  {journal}
  {\bibinfo  {journal} {Rep. Prog. Phys.}\ }\textbf {\bibinfo {volume} {77}},\
  \bibinfo {pages} {034401} (\bibinfo {year} {2014})}\BibitemShut {NoStop}%
\bibitem [{\citenamefont {Schmidt}\ \emph {et~al.}(2018)\citenamefont
  {Schmidt}, \citenamefont {Knap}, \citenamefont {Ivanov}, \citenamefont {You},
  \citenamefont {Cetina},\ and\ \citenamefont {Demler}}]{schmidt2018universal}%
  \BibitemOpen
  \bibfield  {author} {\bibinfo {author} {\bibfnamefont {R.}~\bibnamefont
  {Schmidt}}, \bibinfo {author} {\bibfnamefont {M.}~\bibnamefont {Knap}},
  \bibinfo {author} {\bibfnamefont {D.~A.}\ \bibnamefont {Ivanov}}, \bibinfo
  {author} {\bibfnamefont {J.-S.}\ \bibnamefont {You}}, \bibinfo {author}
  {\bibfnamefont {M.}~\bibnamefont {Cetina}},\ and\ \bibinfo {author}
  {\bibfnamefont {E.}~\bibnamefont {Demler}},\ }\bibfield  {title} {\bibinfo
  {title} {Universal many-body response of heavy impurities coupled to a fermi
  sea: a review of recent progress},\ }\href
  {https://doi.org/10.1088/1361-6633/aa9593} {\bibfield  {journal} {\bibinfo
  {journal} {Rep. Prog. Phys.}\ }\textbf {\bibinfo {volume} {81}},\ \bibinfo
  {pages} {024401} (\bibinfo {year} {2018})}\BibitemShut {NoStop}%
\bibitem [{\citenamefont {Nascimbene}\ \emph {et~al.}(2009)\citenamefont
  {Nascimbene}, \citenamefont {Navon}, \citenamefont {Jiang}, \citenamefont
  {Tarruell}, \citenamefont {Teichmann}, \citenamefont {Mckeever},
  \citenamefont {Chevy},\ and\ \citenamefont
  {Salomon}}]{nascimbene2009collective}%
  \BibitemOpen
  \bibfield  {author} {\bibinfo {author} {\bibfnamefont {S.}~\bibnamefont
  {Nascimbene}}, \bibinfo {author} {\bibfnamefont {N.}~\bibnamefont {Navon}},
  \bibinfo {author} {\bibfnamefont {K.~J.}\ \bibnamefont {Jiang}}, \bibinfo
  {author} {\bibfnamefont {L.}~\bibnamefont {Tarruell}}, \bibinfo {author}
  {\bibfnamefont {M.}~\bibnamefont {Teichmann}}, \bibinfo {author}
  {\bibfnamefont {J.}~\bibnamefont {Mckeever}}, \bibinfo {author}
  {\bibfnamefont {F.}~\bibnamefont {Chevy}},\ and\ \bibinfo {author}
  {\bibfnamefont {C.}~\bibnamefont {Salomon}},\ }\bibfield  {title} {\bibinfo
  {title} {Collective oscillations of an imbalanced fermi gas: axial
  compression modes and polaron effective mass},\ }\href
  {https://doi.org/10.1103/PhysRevLett.103.170402} {\bibfield  {journal}
  {\bibinfo  {journal} {Phys. Rev. Lett.}\ }\textbf {\bibinfo {volume} {103}},\
  \bibinfo {pages} {170402} (\bibinfo {year} {2009})}\BibitemShut {NoStop}%
\bibitem [{\citenamefont {Schirotzek}\ \emph {et~al.}(2009)\citenamefont
  {Schirotzek}, \citenamefont {Wu}, \citenamefont {Sommer},\ and\ \citenamefont
  {Zwierlein}}]{schirotzek2009observation}%
  \BibitemOpen
  \bibfield  {author} {\bibinfo {author} {\bibfnamefont {A.}~\bibnamefont
  {Schirotzek}}, \bibinfo {author} {\bibfnamefont {C.-H.}\ \bibnamefont {Wu}},
  \bibinfo {author} {\bibfnamefont {A.}~\bibnamefont {Sommer}},\ and\ \bibinfo
  {author} {\bibfnamefont {M.~W.}\ \bibnamefont {Zwierlein}},\ }\bibfield
  {title} {\bibinfo {title} {Observation of fermi polarons in a tunable fermi
  liquid of ultracold atoms},\ }\href
  {https://doi.org/10.1103/PhysRevLett.102.230402} {\bibfield  {journal}
  {\bibinfo  {journal} {Phys. Rev. Lett.}\ }\textbf {\bibinfo {volume} {102}},\
  \bibinfo {pages} {230402} (\bibinfo {year} {2009})}\BibitemShut {NoStop}%
\bibitem [{\citenamefont {Palzer}\ \emph {et~al.}(2009)\citenamefont {Palzer},
  \citenamefont {Zipkes}, \citenamefont {Sias},\ and\ \citenamefont
  {K\"ohl}}]{palzer_impurity_transport_09}%
  \BibitemOpen
  \bibfield  {author} {\bibinfo {author} {\bibfnamefont {S.}~\bibnamefont
  {Palzer}}, \bibinfo {author} {\bibfnamefont {C.}~\bibnamefont {Zipkes}},
  \bibinfo {author} {\bibfnamefont {C.}~\bibnamefont {Sias}},\ and\ \bibinfo
  {author} {\bibfnamefont {M.}~\bibnamefont {K\"ohl}},\ }\bibfield  {title}
  {\bibinfo {title} {Quantum transport through a tonks-girardeau gas},\ }\href
  {https://doi.org/10.1103/PhysRevLett.103.150601} {\bibfield  {journal}
  {\bibinfo  {journal} {Phys. Rev. Lett.}\ }\textbf {\bibinfo {volume} {103}},\
  \bibinfo {pages} {150601} (\bibinfo {year} {2009})},\ \Eprint
  {https://arxiv.org/abs/arXiv:0903.4823} {arXiv:0903.4823} \BibitemShut
  {NoStop}%
\bibitem [{\citenamefont {Catani}\ \emph {et~al.}(2012)\citenamefont {Catani},
  \citenamefont {Lamporesi}, \citenamefont {Naik}, \citenamefont {Gring},
  \citenamefont {Inguscio}, \citenamefont {Minardi}, \citenamefont {Kantian},\
  and\ \citenamefont {Giamarchi}}]{catani_impurity_dynamics_11}%
  \BibitemOpen
  \bibfield  {author} {\bibinfo {author} {\bibfnamefont {J.}~\bibnamefont
  {Catani}}, \bibinfo {author} {\bibfnamefont {G.}~\bibnamefont {Lamporesi}},
  \bibinfo {author} {\bibfnamefont {D.}~\bibnamefont {Naik}}, \bibinfo {author}
  {\bibfnamefont {M.}~\bibnamefont {Gring}}, \bibinfo {author} {\bibfnamefont
  {M.}~\bibnamefont {Inguscio}}, \bibinfo {author} {\bibfnamefont
  {F.}~\bibnamefont {Minardi}}, \bibinfo {author} {\bibfnamefont
  {A.}~\bibnamefont {Kantian}},\ and\ \bibinfo {author} {\bibfnamefont
  {T.}~\bibnamefont {Giamarchi}},\ }\bibfield  {title} {\bibinfo {title}
  {Quantum dynamics of impurities in a one-dimensional {Bose} gas},\ }\href
  {https://doi.org/10.1103/PhysRevA.85.023623} {\bibfield  {journal} {\bibinfo
  {journal} {Phys. Rev. A}\ }\textbf {\bibinfo {volume} {85}},\ \bibinfo
  {pages} {023623} (\bibinfo {year} {2012})},\ \Eprint
  {https://arxiv.org/abs/arXiv:1106.0828} {arXiv:1106.0828} \BibitemShut
  {NoStop}%
\bibitem [{\citenamefont {Kohstall}\ \emph {et~al.}(2012)\citenamefont
  {Kohstall}, \citenamefont {Zaccanti}, \citenamefont {Jag}, \citenamefont
  {Trenkwalder}, \citenamefont {Massignan}, \citenamefont {Bruun},
  \citenamefont {Schreck},\ and\ \citenamefont
  {Grimm}}]{kohstall2012metastability}%
  \BibitemOpen
  \bibfield  {author} {\bibinfo {author} {\bibfnamefont {C.}~\bibnamefont
  {Kohstall}}, \bibinfo {author} {\bibfnamefont {M.}~\bibnamefont {Zaccanti}},
  \bibinfo {author} {\bibfnamefont {M.}~\bibnamefont {Jag}}, \bibinfo {author}
  {\bibfnamefont {A.}~\bibnamefont {Trenkwalder}}, \bibinfo {author}
  {\bibfnamefont {P.}~\bibnamefont {Massignan}}, \bibinfo {author}
  {\bibfnamefont {G.~M.}\ \bibnamefont {Bruun}}, \bibinfo {author}
  {\bibfnamefont {F.}~\bibnamefont {Schreck}},\ and\ \bibinfo {author}
  {\bibfnamefont {R.}~\bibnamefont {Grimm}},\ }\bibfield  {title} {\bibinfo
  {title} {Metastability and coherence of repulsive polarons in a strongly
  interacting fermi mixture},\ }\href {https://doi.org/10.1038/nature11065}
  {\bibfield  {journal} {\bibinfo  {journal} {Nature}\ }\textbf {\bibinfo
  {volume} {485}},\ \bibinfo {pages} {615} (\bibinfo {year}
  {2012})}\BibitemShut {NoStop}%
\bibitem [{\citenamefont {Stojanovi{\'c}}\ \emph {et~al.}(2012)\citenamefont
  {Stojanovi{\'c}}, \citenamefont {Shi}, \citenamefont {Bruder},\ and\
  \citenamefont {Cirac}}]{stojanovic2012quantum}%
  \BibitemOpen
  \bibfield  {author} {\bibinfo {author} {\bibfnamefont {V.~M.}\ \bibnamefont
  {Stojanovi{\'c}}}, \bibinfo {author} {\bibfnamefont {T.}~\bibnamefont {Shi}},
  \bibinfo {author} {\bibfnamefont {C.}~\bibnamefont {Bruder}},\ and\ \bibinfo
  {author} {\bibfnamefont {J.~I.}\ \bibnamefont {Cirac}},\ }\bibfield  {title}
  {\bibinfo {title} {Quantum simulation of small-polaron formation with trapped
  ions},\ }\href {https://doi.org/10.1103/PhysRevLett.109.250501} {\bibfield
  {journal} {\bibinfo  {journal} {Phys. Rev. Lett.}\ }\textbf {\bibinfo
  {volume} {109}},\ \bibinfo {pages} {250501} (\bibinfo {year}
  {2012})}\BibitemShut {NoStop}%
\bibitem [{\citenamefont {Fukuhara}\ \emph {et~al.}(2013)\citenamefont
  {Fukuhara}, \citenamefont {Kantian}, \citenamefont {Endres}, \citenamefont
  {Cheneau}, \citenamefont {Schau\ss}, \citenamefont {Hild}, \citenamefont
  {Bellem}, \citenamefont {Schollw\"ock}, \citenamefont {Giamarchi},
  \citenamefont {Gross}, \citenamefont {Bloch},\ and\ \citenamefont
  {Kuhr}}]{fukuhara_spin_impurity_2013}%
  \BibitemOpen
  \bibfield  {author} {\bibinfo {author} {\bibfnamefont {T.}~\bibnamefont
  {Fukuhara}}, \bibinfo {author} {\bibfnamefont {A.}~\bibnamefont {Kantian}},
  \bibinfo {author} {\bibfnamefont {M.}~\bibnamefont {Endres}}, \bibinfo
  {author} {\bibfnamefont {M.}~\bibnamefont {Cheneau}}, \bibinfo {author}
  {\bibfnamefont {P.}~\bibnamefont {Schau\ss}}, \bibinfo {author}
  {\bibfnamefont {S.}~\bibnamefont {Hild}}, \bibinfo {author} {\bibfnamefont
  {D.}~\bibnamefont {Bellem}}, \bibinfo {author} {\bibfnamefont
  {U.}~\bibnamefont {Schollw\"ock}}, \bibinfo {author} {\bibfnamefont
  {T.}~\bibnamefont {Giamarchi}}, \bibinfo {author} {\bibfnamefont
  {C.}~\bibnamefont {Gross}}, \bibinfo {author} {\bibfnamefont
  {I.}~\bibnamefont {Bloch}},\ and\ \bibinfo {author} {\bibfnamefont
  {S.}~\bibnamefont {Kuhr}},\ }\bibfield  {title} {\bibinfo {title} {Quantum
  dynamics of a mobile spin impurity},\ }\href
  {https://doi.org/doi:10.1038/nphys2561} {\bibfield  {journal} {\bibinfo
  {journal} {Nat. Phys.}\ }\textbf {\bibinfo {volume} {9}},\ \bibinfo {pages}
  {235} (\bibinfo {year} {2013})},\ \Eprint
  {https://arxiv.org/abs/arXiv:1209.6468} {arXiv:1209.6468} \BibitemShut
  {NoStop}%
\bibitem [{\citenamefont {Hu}\ \emph {et~al.}(2016)\citenamefont {Hu},
  \citenamefont {Van~de Graaff}, \citenamefont {Kedar}, \citenamefont {Corson},
  \citenamefont {Cornell},\ and\ \citenamefont {Jin}}]{hu2016bose}%
  \BibitemOpen
  \bibfield  {author} {\bibinfo {author} {\bibfnamefont {M.-G.}\ \bibnamefont
  {Hu}}, \bibinfo {author} {\bibfnamefont {M.~J.}\ \bibnamefont {Van~de
  Graaff}}, \bibinfo {author} {\bibfnamefont {D.}~\bibnamefont {Kedar}},
  \bibinfo {author} {\bibfnamefont {J.~P.}\ \bibnamefont {Corson}}, \bibinfo
  {author} {\bibfnamefont {E.~A.}\ \bibnamefont {Cornell}},\ and\ \bibinfo
  {author} {\bibfnamefont {D.~S.}\ \bibnamefont {Jin}},\ }\bibfield  {title}
  {\bibinfo {title} {Bose polarons in the strongly interacting regime},\ }\href
  {https://doi.org/10.1103/PhysRevLett.117.055301} {\bibfield  {journal}
  {\bibinfo  {journal} {Phys. Rev. Lett.}\ }\textbf {\bibinfo {volume} {117}},\
  \bibinfo {pages} {055301} (\bibinfo {year} {2016})}\BibitemShut {NoStop}%
\bibitem [{\citenamefont {J{\o}rgensen}\ \emph {et~al.}(2016)\citenamefont
  {J{\o}rgensen}, \citenamefont {Wacker}, \citenamefont {Skalmstang},
  \citenamefont {Parish}, \citenamefont {Levinsen}, \citenamefont
  {Christensen}, \citenamefont {Bruun},\ and\ \citenamefont
  {Arlt}}]{jorgensen2016observation}%
  \BibitemOpen
  \bibfield  {author} {\bibinfo {author} {\bibfnamefont {N.~B.}\ \bibnamefont
  {J{\o}rgensen}}, \bibinfo {author} {\bibfnamefont {L.}~\bibnamefont
  {Wacker}}, \bibinfo {author} {\bibfnamefont {K.~T.}\ \bibnamefont
  {Skalmstang}}, \bibinfo {author} {\bibfnamefont {M.~M.}\ \bibnamefont
  {Parish}}, \bibinfo {author} {\bibfnamefont {J.}~\bibnamefont {Levinsen}},
  \bibinfo {author} {\bibfnamefont {R.~S.}\ \bibnamefont {Christensen}},
  \bibinfo {author} {\bibfnamefont {G.~M.}\ \bibnamefont {Bruun}},\ and\
  \bibinfo {author} {\bibfnamefont {J.~J.}\ \bibnamefont {Arlt}},\ }\bibfield
  {title} {\bibinfo {title} {Observation of attractive and repulsive polarons
  in a bose-einstein condensate},\ }\href
  {https://doi.org/10.1103/PhysRevLett.117.055302} {\bibfield  {journal}
  {\bibinfo  {journal} {Phys. Rev. Lett.}\ }\textbf {\bibinfo {volume} {117}},\
  \bibinfo {pages} {055302} (\bibinfo {year} {2016})}\BibitemShut {NoStop}%
\bibitem [{\citenamefont {Cetina}\ \emph {et~al.}(2016)\citenamefont {Cetina},
  \citenamefont {Jag}, \citenamefont {Lous}, \citenamefont {Fritsche},
  \citenamefont {Walraven}, \citenamefont {Grimm}, \citenamefont {Levinsen},
  \citenamefont {Parish}, \citenamefont {Schmidt}, \citenamefont {Knap},\ and\
  \citenamefont {et~al.}}]{cetina2016ultrafast}%
  \BibitemOpen
  \bibfield  {author} {\bibinfo {author} {\bibfnamefont {M.}~\bibnamefont
  {Cetina}}, \bibinfo {author} {\bibfnamefont {M.}~\bibnamefont {Jag}},
  \bibinfo {author} {\bibfnamefont {R.~S.}\ \bibnamefont {Lous}}, \bibinfo
  {author} {\bibfnamefont {I.}~\bibnamefont {Fritsche}}, \bibinfo {author}
  {\bibfnamefont {J.~T.~M.}\ \bibnamefont {Walraven}}, \bibinfo {author}
  {\bibfnamefont {R.}~\bibnamefont {Grimm}}, \bibinfo {author} {\bibfnamefont
  {J.}~\bibnamefont {Levinsen}}, \bibinfo {author} {\bibfnamefont {M.~M.}\
  \bibnamefont {Parish}}, \bibinfo {author} {\bibfnamefont {R.}~\bibnamefont
  {Schmidt}}, \bibinfo {author} {\bibfnamefont {M.}~\bibnamefont {Knap}},\ and\
  \bibinfo {author} {\bibnamefont {et~al.}},\ }\bibfield  {title} {\bibinfo
  {title} {Ultrafast many-body interferometry of impurities coupled to a fermi
  sea},\ }\href {https://doi.org/10.1126/science.aaf5134} {\bibfield  {journal}
  {\bibinfo  {journal} {Science}\ }\textbf {\bibinfo {volume} {354}},\ \bibinfo
  {pages} {96} (\bibinfo {year} {2016})}\BibitemShut {NoStop}%
\bibitem [{\citenamefont {Meinert}\ \emph {et~al.}(2017)\citenamefont
  {Meinert}, \citenamefont {Knap}, \citenamefont {Kirilov}, \citenamefont
  {Jag-Lauber}, \citenamefont {Zvonarev}, \citenamefont {Demler},\ and\
  \citenamefont {N{\"a}gerl}}]{meinert2017bloch}%
  \BibitemOpen
  \bibfield  {author} {\bibinfo {author} {\bibfnamefont {F.}~\bibnamefont
  {Meinert}}, \bibinfo {author} {\bibfnamefont {M.}~\bibnamefont {Knap}},
  \bibinfo {author} {\bibfnamefont {E.}~\bibnamefont {Kirilov}}, \bibinfo
  {author} {\bibfnamefont {K.}~\bibnamefont {Jag-Lauber}}, \bibinfo {author}
  {\bibfnamefont {M.~B.}\ \bibnamefont {Zvonarev}}, \bibinfo {author}
  {\bibfnamefont {E.}~\bibnamefont {Demler}},\ and\ \bibinfo {author}
  {\bibfnamefont {H.-C.}\ \bibnamefont {N{\"a}gerl}},\ }\bibfield  {title}
  {\bibinfo {title} {Bloch oscillations in the absence of a lattice},\ }\href
  {https://doi.org/10.1126/science.aah6616} {\bibfield  {journal} {\bibinfo
  {journal} {Science}\ }\textbf {\bibinfo {volume} {356}},\ \bibinfo {pages}
  {945} (\bibinfo {year} {2017})},\ \Eprint
  {https://arxiv.org/abs/arXiv:1608.08200} {arXiv:1608.08200} \BibitemShut
  {NoStop}%
\bibitem [{\citenamefont {Yan}\ \emph {et~al.}(2019)\citenamefont {Yan},
  \citenamefont {Patel}, \citenamefont {Mukherjee}, \citenamefont {Fletcher},
  \citenamefont {Struck},\ and\ \citenamefont {Zwierlein}}]{yan2019boiling}%
  \BibitemOpen
  \bibfield  {author} {\bibinfo {author} {\bibfnamefont {Z.}~\bibnamefont
  {Yan}}, \bibinfo {author} {\bibfnamefont {P.~B.}\ \bibnamefont {Patel}},
  \bibinfo {author} {\bibfnamefont {B.}~\bibnamefont {Mukherjee}}, \bibinfo
  {author} {\bibfnamefont {R.~J.}\ \bibnamefont {Fletcher}}, \bibinfo {author}
  {\bibfnamefont {J.}~\bibnamefont {Struck}},\ and\ \bibinfo {author}
  {\bibfnamefont {M.~W.}\ \bibnamefont {Zwierlein}},\ }\bibfield  {title}
  {\bibinfo {title} {Boiling a unitary fermi liquid},\ }\href
  {https://doi.org/10.1103/PhysRevLett.122.093401} {\bibfield  {journal}
  {\bibinfo  {journal} {Phys. Rev. Lett.}\ }\textbf {\bibinfo {volume} {122}},\
  \bibinfo {pages} {093401} (\bibinfo {year} {2019})}\BibitemShut {NoStop}%
\bibitem [{\citenamefont {Yan}\ \emph {et~al.}(2020)\citenamefont {Yan},
  \citenamefont {Ni}, \citenamefont {Robens},\ and\ \citenamefont
  {Zwierlein}}]{yan2020bose}%
  \BibitemOpen
  \bibfield  {author} {\bibinfo {author} {\bibfnamefont {Z.~Z.}\ \bibnamefont
  {Yan}}, \bibinfo {author} {\bibfnamefont {Y.}~\bibnamefont {Ni}}, \bibinfo
  {author} {\bibfnamefont {C.}~\bibnamefont {Robens}},\ and\ \bibinfo {author}
  {\bibfnamefont {M.~W.}\ \bibnamefont {Zwierlein}},\ }\bibfield  {title}
  {\bibinfo {title} {Bose polarons near quantum criticality},\ }\href
  {https://doi.org/10.1126/science.aax5850} {\bibfield  {journal} {\bibinfo
  {journal} {Science}\ }\textbf {\bibinfo {volume} {368}},\ \bibinfo {pages}
  {190} (\bibinfo {year} {2020})}\BibitemShut {NoStop}%
\bibitem [{\citenamefont {Bloch}\ \emph {et~al.}(2008)\citenamefont {Bloch},
  \citenamefont {Dalibard},\ and\ \citenamefont {Zwerger}}]{bloch2008many}%
  \BibitemOpen
  \bibfield  {author} {\bibinfo {author} {\bibfnamefont {I.}~\bibnamefont
  {Bloch}}, \bibinfo {author} {\bibfnamefont {J.}~\bibnamefont {Dalibard}},\
  and\ \bibinfo {author} {\bibfnamefont {W.}~\bibnamefont {Zwerger}},\
  }\bibfield  {title} {\bibinfo {title} {Many-body physics with ultracold
  gases},\ }\href {https://doi.org/10.1103/RevModPhys.80.885} {\bibfield
  {journal} {\bibinfo  {journal} {Rev. Mod. Phys.}\ }\textbf {\bibinfo {volume}
  {80}},\ \bibinfo {pages} {885} (\bibinfo {year} {2008})},\ \Eprint
  {https://arxiv.org/abs/arXiv:0704.3011} {arXiv:0704.3011} \BibitemShut
  {NoStop}%
\bibitem [{\citenamefont {Chevy}(2006)}]{chevy2006universal}%
  \BibitemOpen
  \bibfield  {author} {\bibinfo {author} {\bibfnamefont {F.}~\bibnamefont
  {Chevy}},\ }\bibfield  {title} {\bibinfo {title} {Universal phase diagram of
  a strongly interacting fermi gas with unbalanced spin populations},\ }\href
  {https://doi.org/10.1103/PhysRevA.74.063628} {\bibfield  {journal} {\bibinfo
  {journal} {Phys. Rev. A}\ }\textbf {\bibinfo {volume} {74}},\ \bibinfo
  {pages} {063628} (\bibinfo {year} {2006})}\BibitemShut {NoStop}%
\bibitem [{\citenamefont {Parish}\ and\ \citenamefont
  {Levinsen}(2013)}]{parish2013highly}%
  \BibitemOpen
  \bibfield  {author} {\bibinfo {author} {\bibfnamefont {M.~M.}\ \bibnamefont
  {Parish}}\ and\ \bibinfo {author} {\bibfnamefont {J.}~\bibnamefont
  {Levinsen}},\ }\bibfield  {title} {\bibinfo {title} {Highly polarized fermi
  gases in two dimensions},\ }\href
  {https://doi.org/10.1103/PhysRevA.87.033616} {\bibfield  {journal} {\bibinfo
  {journal} {Phys. Rev. A}\ }\textbf {\bibinfo {volume} {87}},\ \bibinfo
  {pages} {033616} (\bibinfo {year} {2013})}\BibitemShut {NoStop}%
\bibitem [{\citenamefont {Peotta}\ \emph {et~al.}(2013)\citenamefont {Peotta},
  \citenamefont {Rossini}, \citenamefont {Polini}, \citenamefont {Minardi},\
  and\ \citenamefont {Fazio}}]{peotta_mobile_impurity_TDMRG_2013}%
  \BibitemOpen
  \bibfield  {author} {\bibinfo {author} {\bibfnamefont {S.}~\bibnamefont
  {Peotta}}, \bibinfo {author} {\bibfnamefont {D.}~\bibnamefont {Rossini}},
  \bibinfo {author} {\bibfnamefont {M.}~\bibnamefont {Polini}}, \bibinfo
  {author} {\bibfnamefont {F.}~\bibnamefont {Minardi}},\ and\ \bibinfo {author}
  {\bibfnamefont {R.}~\bibnamefont {Fazio}},\ }\bibfield  {title} {\bibinfo
  {title} {Quantum breathing of an impurity in a one-dimensional bath of
  interacting bosons},\ }\href {https://doi.org/10.1103/PhysRevLett.110.015302}
  {\bibfield  {journal} {\bibinfo  {journal} {Phys. Rev. Lett.}\ }\textbf
  {\bibinfo {volume} {110}},\ \bibinfo {pages} {015302} (\bibinfo {year}
  {2013})},\ \Eprint {https://arxiv.org/abs/arXiv:1206.3984} {arXiv:1206.3984}
  \BibitemShut {NoStop}%
\bibitem [{\citenamefont {Massel}\ \emph {et~al.}(2013)\citenamefont {Massel},
  \citenamefont {Kantian}, \citenamefont {Daley}, \citenamefont {Giamarchi},\
  and\ \citenamefont {T{\"o}rm{\"a}}}]{massel_mobile_impurity_TDMRG_2013}%
  \BibitemOpen
  \bibfield  {author} {\bibinfo {author} {\bibfnamefont {F.}~\bibnamefont
  {Massel}}, \bibinfo {author} {\bibfnamefont {A.}~\bibnamefont {Kantian}},
  \bibinfo {author} {\bibfnamefont {A.~J.}\ \bibnamefont {Daley}}, \bibinfo
  {author} {\bibfnamefont {T.}~\bibnamefont {Giamarchi}},\ and\ \bibinfo
  {author} {\bibfnamefont {P.}~\bibnamefont {T{\"o}rm{\"a}}},\ }\bibfield
  {title} {\bibinfo {title} {Dynamics of an impurity in a one-dimensional
  lattice},\ }\href {https://doi.org/10.1088/1367-2630/15/4/045018} {\bibfield
  {journal} {\bibinfo  {journal} {New J. Phys.}\ }\textbf {\bibinfo {volume}
  {15}},\ \bibinfo {pages} {045018} (\bibinfo {year} {2013})},\ \Eprint
  {https://arxiv.org/abs/arXiv:1210.4270} {arXiv:1210.4270} \BibitemShut
  {NoStop}%
\bibitem [{\citenamefont {Knap}\ \emph {et~al.}(2014)\citenamefont {Knap},
  \citenamefont {Mathy}, \citenamefont {Ganahl}, \citenamefont {Zvonarev},\
  and\ \citenamefont {Demler}}]{knap2014quantum}%
  \BibitemOpen
  \bibfield  {author} {\bibinfo {author} {\bibfnamefont {M.}~\bibnamefont
  {Knap}}, \bibinfo {author} {\bibfnamefont {C.~J.~M.}\ \bibnamefont {Mathy}},
  \bibinfo {author} {\bibfnamefont {M.}~\bibnamefont {Ganahl}}, \bibinfo
  {author} {\bibfnamefont {M.~B.}\ \bibnamefont {Zvonarev}},\ and\ \bibinfo
  {author} {\bibfnamefont {E.}~\bibnamefont {Demler}},\ }\bibfield  {title}
  {\bibinfo {title} {Quantum flutter: Signatures and robustness},\ }\href
  {https://doi.org/10.1103/PhysRevLett.112.015302} {\bibfield  {journal}
  {\bibinfo  {journal} {Phys. Rev. Lett.}\ }\textbf {\bibinfo {volume} {112}},\
  \bibinfo {pages} {015302} (\bibinfo {year} {2014})},\ \Eprint
  {https://arxiv.org/abs/arXiv:1303.3583} {arXiv:1303.3583} \BibitemShut
  {NoStop}%
\bibitem [{\citenamefont {Li}\ and\ \citenamefont
  {Sarma}(2014)}]{li2014variational}%
  \BibitemOpen
  \bibfield  {author} {\bibinfo {author} {\bibfnamefont {W.}~\bibnamefont
  {Li}}\ and\ \bibinfo {author} {\bibfnamefont {S.~D.}\ \bibnamefont {Sarma}},\
  }\bibfield  {title} {\bibinfo {title} {Variational study of polarons in
  bose-einstein condensates},\ }\href
  {https://doi.org/10.1103/PhysRevA.90.013618} {\bibfield  {journal} {\bibinfo
  {journal} {Phys. Rev. A}\ }\textbf {\bibinfo {volume} {90}},\ \bibinfo
  {pages} {013618} (\bibinfo {year} {2014})}\BibitemShut {NoStop}%
\bibitem [{\citenamefont {Shchadilova}\ \emph {et~al.}(2016)\citenamefont
  {Shchadilova}, \citenamefont {Schmidt}, \citenamefont {Grusdt},\ and\
  \citenamefont {Demler}}]{shchadilova2016quantum}%
  \BibitemOpen
  \bibfield  {author} {\bibinfo {author} {\bibfnamefont {Y.~E.}\ \bibnamefont
  {Shchadilova}}, \bibinfo {author} {\bibfnamefont {R.}~\bibnamefont
  {Schmidt}}, \bibinfo {author} {\bibfnamefont {F.}~\bibnamefont {Grusdt}},\
  and\ \bibinfo {author} {\bibfnamefont {E.}~\bibnamefont {Demler}},\
  }\bibfield  {title} {\bibinfo {title} {Quantum dynamics of ultracold bose
  polarons},\ }\href {https://doi.org/10.1103/PhysRevLett.117.113002}
  {\bibfield  {journal} {\bibinfo  {journal} {Phys. Rev. Lett.}\ }\textbf
  {\bibinfo {volume} {117}},\ \bibinfo {pages} {113002} (\bibinfo {year}
  {2016})}\BibitemShut {NoStop}%
\bibitem [{\citenamefont {Kain}\ and\ \citenamefont
  {Ling}(2017)}]{kain2017hartree}%
  \BibitemOpen
  \bibfield  {author} {\bibinfo {author} {\bibfnamefont {B.}~\bibnamefont
  {Kain}}\ and\ \bibinfo {author} {\bibfnamefont {H.~Y.}\ \bibnamefont
  {Ling}},\ }\bibfield  {title} {\bibinfo {title} {Hartree-fock treatment of
  fermi polarons using the lee-low-pines transformation},\ }\href
  {https://doi.org/10.1103/PhysRevA.96.033627} {\bibfield  {journal} {\bibinfo
  {journal} {Phys. Rev. A}\ }\textbf {\bibinfo {volume} {96}},\ \bibinfo
  {pages} {033627} (\bibinfo {year} {2017})}\BibitemShut {NoStop}%
\bibitem [{\citenamefont {Shi}\ \emph {et~al.}(2018)\citenamefont {Shi},
  \citenamefont {Demler},\ and\ \citenamefont {Cirac}}]{shi2018variational}%
  \BibitemOpen
  \bibfield  {author} {\bibinfo {author} {\bibfnamefont {T.}~\bibnamefont
  {Shi}}, \bibinfo {author} {\bibfnamefont {E.}~\bibnamefont {Demler}},\ and\
  \bibinfo {author} {\bibfnamefont {J.~I.}\ \bibnamefont {Cirac}},\ }\bibfield
  {title} {\bibinfo {title} {Variational study of fermionic and bosonic systems
  with non-gaussian states: Theory and applications},\ }\href
  {https://doi.org/10.1016/j.aop.2017.11.014} {\bibfield  {journal} {\bibinfo
  {journal} {Ann. Phys.}\ }\textbf {\bibinfo {volume} {390}},\ \bibinfo {pages}
  {245} (\bibinfo {year} {2018})}\BibitemShut {NoStop}%
\bibitem [{\citenamefont {Mistakidis}\ \emph {et~al.}(2019)\citenamefont
  {Mistakidis}, \citenamefont {Katsimiga}, \citenamefont {Koutentakis},\ and\
  \citenamefont {Schmelcher}}]{mistakidis2019repulsive}%
  \BibitemOpen
  \bibfield  {author} {\bibinfo {author} {\bibfnamefont {S.~I.}\ \bibnamefont
  {Mistakidis}}, \bibinfo {author} {\bibfnamefont {G.~C.}\ \bibnamefont
  {Katsimiga}}, \bibinfo {author} {\bibfnamefont {G.~M.}\ \bibnamefont
  {Koutentakis}},\ and\ \bibinfo {author} {\bibfnamefont {P.}~\bibnamefont
  {Schmelcher}},\ }\bibfield  {title} {\bibinfo {title} {Repulsive fermi
  polarons and their induced interactions in binary mixtures of ultracold
  atoms},\ }\href {https://doi.org/10.1088/1367-2630/ab1045} {\bibfield
  {journal} {\bibinfo  {journal} {New J. Phys.}\ }\textbf {\bibinfo {volume}
  {21}},\ \bibinfo {pages} {043032} (\bibinfo {year} {2019})}\BibitemShut
  {NoStop}%
\bibitem [{\citenamefont {Grusdt}\ \emph {et~al.}(2015)\citenamefont {Grusdt},
  \citenamefont {Shchadilova}, \citenamefont {Rubtsov},\ and\ \citenamefont
  {Demler}}]{grusdt2015renormalization}%
  \BibitemOpen
  \bibfield  {author} {\bibinfo {author} {\bibfnamefont {F.}~\bibnamefont
  {Grusdt}}, \bibinfo {author} {\bibfnamefont {Y.~E.}\ \bibnamefont
  {Shchadilova}}, \bibinfo {author} {\bibfnamefont {A.~N.}\ \bibnamefont
  {Rubtsov}},\ and\ \bibinfo {author} {\bibfnamefont {E.}~\bibnamefont
  {Demler}},\ }\bibfield  {title} {\bibinfo {title} {Renormalization group
  approach to the fr{\"o}hlich polaron model: application to impurity-bec
  problem},\ }\href {https://doi.org/10.1038/srep12124} {\bibfield  {journal}
  {\bibinfo  {journal} {Sci. Rep.}\ }\textbf {\bibinfo {volume} {5}},\ \bibinfo
  {pages} {12124} (\bibinfo {year} {2015})}\BibitemShut {NoStop}%
\bibitem [{\citenamefont {Grusdt}\ \emph {et~al.}(2018)\citenamefont {Grusdt},
  \citenamefont {Seetharam}, \citenamefont {Shchadilova},\ and\ \citenamefont
  {Demler}}]{grusdt2018strong}%
  \BibitemOpen
  \bibfield  {author} {\bibinfo {author} {\bibfnamefont {F.}~\bibnamefont
  {Grusdt}}, \bibinfo {author} {\bibfnamefont {K.}~\bibnamefont {Seetharam}},
  \bibinfo {author} {\bibfnamefont {Y.}~\bibnamefont {Shchadilova}},\ and\
  \bibinfo {author} {\bibfnamefont {E.}~\bibnamefont {Demler}},\ }\bibfield
  {title} {\bibinfo {title} {Strong-coupling bose polarons out of equilibrium:
  Dynamical renormalization-group approach},\ }\href
  {https://doi.org/10.1103/PhysRevA.97.033612} {\bibfield  {journal} {\bibinfo
  {journal} {Phys. Rev. A}\ }\textbf {\bibinfo {volume} {97}},\ \bibinfo
  {pages} {033612} (\bibinfo {year} {2018})}\BibitemShut {NoStop}%
\bibitem [{\citenamefont {Prokof'ev}\ and\ \citenamefont
  {Svistunov}(2008)}]{prokof2008fermi}%
  \BibitemOpen
  \bibfield  {author} {\bibinfo {author} {\bibfnamefont {N.}~\bibnamefont
  {Prokof'ev}}\ and\ \bibinfo {author} {\bibfnamefont {B.}~\bibnamefont
  {Svistunov}},\ }\bibfield  {title} {\bibinfo {title} {Fermi-polaron problem:
  Diagrammatic monte carlo method for divergent sign-alternating series},\
  }\href {https://doi.org/10.1103/PhysRevB.77.020408} {\bibfield  {journal}
  {\bibinfo  {journal} {Phys. Rev. B}\ }\textbf {\bibinfo {volume} {77}},\
  \bibinfo {pages} {020408} (\bibinfo {year} {2008})},\ \Eprint
  {https://arxiv.org/abs/arXiv:0707.4259} {arXiv:0707.4259} \BibitemShut
  {NoStop}%
\bibitem [{\citenamefont {Ardila}\ and\ \citenamefont
  {Giorgini}(2015)}]{ardila2015impurity}%
  \BibitemOpen
  \bibfield  {author} {\bibinfo {author} {\bibfnamefont {L.~A.~P.}\
  \bibnamefont {Ardila}}\ and\ \bibinfo {author} {\bibfnamefont
  {S.}~\bibnamefont {Giorgini}},\ }\bibfield  {title} {\bibinfo {title}
  {Impurity in a bose-einstein condensate: Study of the attractive and
  repulsive branch using quantum monte carlo methods},\ }\href
  {https://doi.org/10.1103/PhysRevA.92.033612} {\bibfield  {journal} {\bibinfo
  {journal} {Phys. Rev. A}\ }\textbf {\bibinfo {volume} {92}},\ \bibinfo
  {pages} {033612} (\bibinfo {year} {2015})}\BibitemShut {NoStop}%
\bibitem [{\citenamefont {Grusdt}\ \emph {et~al.}(2017)\citenamefont {Grusdt},
  \citenamefont {Astrakharchik},\ and\ \citenamefont
  {Demler}}]{grusdt2017bose}%
  \BibitemOpen
  \bibfield  {author} {\bibinfo {author} {\bibfnamefont {F.}~\bibnamefont
  {Grusdt}}, \bibinfo {author} {\bibfnamefont {G.~E.}\ \bibnamefont
  {Astrakharchik}},\ and\ \bibinfo {author} {\bibfnamefont {E.}~\bibnamefont
  {Demler}},\ }\bibfield  {title} {\bibinfo {title} {Bose polarons in ultracold
  atoms in one dimension: beyond the fr{\"o}hlich paradigm},\ }\href
  {https://doi.org/10.1088/1367-2630/aa8a2e} {\bibfield  {journal} {\bibinfo
  {journal} {New J. Phys.}\ }\textbf {\bibinfo {volume} {19}},\ \bibinfo
  {pages} {103035} (\bibinfo {year} {2017})}\BibitemShut {NoStop}%
\bibitem [{\citenamefont {Rosch}\ and\ \citenamefont
  {Kopp}(1995)}]{rosch1995heavy}%
  \BibitemOpen
  \bibfield  {author} {\bibinfo {author} {\bibfnamefont {A.}~\bibnamefont
  {Rosch}}\ and\ \bibinfo {author} {\bibfnamefont {T.}~\bibnamefont {Kopp}},\
  }\bibfield  {title} {\bibinfo {title} {Heavy particle in a d-dimensional
  fermionic bath: A strong coupling approach},\ }\href
  {https://doi.org/10.1103/PhysRevLett.75.1988} {\bibfield  {journal} {\bibinfo
   {journal} {Phys. Rev. Lett.}\ }\textbf {\bibinfo {volume} {75}},\ \bibinfo
  {pages} {1988} (\bibinfo {year} {1995})}\BibitemShut {NoStop}%
\bibitem [{\citenamefont {Schmidt}\ \emph {et~al.}(2012)\citenamefont
  {Schmidt}, \citenamefont {Enss}, \citenamefont {Pietil{\"a}},\ and\
  \citenamefont {Demler}}]{schmidt2012fermi}%
  \BibitemOpen
  \bibfield  {author} {\bibinfo {author} {\bibfnamefont {R.}~\bibnamefont
  {Schmidt}}, \bibinfo {author} {\bibfnamefont {T.}~\bibnamefont {Enss}},
  \bibinfo {author} {\bibfnamefont {V.}~\bibnamefont {Pietil{\"a}}},\ and\
  \bibinfo {author} {\bibfnamefont {E.}~\bibnamefont {Demler}},\ }\bibfield
  {title} {\bibinfo {title} {Fermi polarons in two dimensions},\ }\href
  {https://doi.org/10.1103/PhysRevA.85.021602} {\bibfield  {journal} {\bibinfo
  {journal} {Phys. Rev. A}\ }\textbf {\bibinfo {volume} {85}},\ \bibinfo
  {pages} {021602} (\bibinfo {year} {2012})}\BibitemShut {NoStop}%
\bibitem [{\citenamefont {Rath}\ and\ \citenamefont
  {Schmidt}(2013)}]{rath2013field}%
  \BibitemOpen
  \bibfield  {author} {\bibinfo {author} {\bibfnamefont {S.~P.}\ \bibnamefont
  {Rath}}\ and\ \bibinfo {author} {\bibfnamefont {R.}~\bibnamefont {Schmidt}},\
  }\bibfield  {title} {\bibinfo {title} {Field-theoretical study of the bose
  polaron},\ }\href {https://doi.org/10.1103/PhysRevA.88.053632} {\bibfield
  {journal} {\bibinfo  {journal} {Phys. Rev. A}\ }\textbf {\bibinfo {volume}
  {88}},\ \bibinfo {pages} {053632} (\bibinfo {year} {2013})}\BibitemShut
  {NoStop}%
\bibitem [{\citenamefont {Burovski}\ \emph {et~al.}(2014)\citenamefont
  {Burovski}, \citenamefont {Cheianov}, \citenamefont {Gamayun},\ and\
  \citenamefont {Lychkovskiy}}]{burovski_impurity_momentum_2014}%
  \BibitemOpen
  \bibfield  {author} {\bibinfo {author} {\bibfnamefont {E.}~\bibnamefont
  {Burovski}}, \bibinfo {author} {\bibfnamefont {V.}~\bibnamefont {Cheianov}},
  \bibinfo {author} {\bibfnamefont {O.}~\bibnamefont {Gamayun}},\ and\ \bibinfo
  {author} {\bibfnamefont {O.}~\bibnamefont {Lychkovskiy}},\ }\bibfield
  {title} {\bibinfo {title} {Momentum relaxation of a mobile impurity in a
  one-dimensional quantum gas},\ }\href
  {https://doi.org/10.1103/PhysRevA.89.041601} {\bibfield  {journal} {\bibinfo
  {journal} {Phys. Rev. A}\ }\textbf {\bibinfo {volume} {89}},\ \bibinfo
  {pages} {041601} (\bibinfo {year} {2014})},\ \Eprint
  {https://arxiv.org/abs/arXiv:1308.6147} {arXiv:1308.6147} \BibitemShut
  {NoStop}%
\bibitem [{\citenamefont {Gamayun}\ \emph {et~al.}(2014)\citenamefont
  {Gamayun}, \citenamefont {Lychkovskiy},\ and\ \citenamefont
  {Cheianov}}]{gamayun_kinetic_impurity_TG_14}%
  \BibitemOpen
  \bibfield  {author} {\bibinfo {author} {\bibfnamefont {O.}~\bibnamefont
  {Gamayun}}, \bibinfo {author} {\bibfnamefont {O.}~\bibnamefont
  {Lychkovskiy}},\ and\ \bibinfo {author} {\bibfnamefont {V.}~\bibnamefont
  {Cheianov}},\ }\bibfield  {title} {\bibinfo {title} {{Kinetic theory for a
  mobile impurity in a degenerate Tonks-Girardeau gas}},\ }\href
  {https://doi.org/10.1103/PhysRevE.90.032132} {\bibfield  {journal} {\bibinfo
  {journal} {Phys. Rev. E}\ }\textbf {\bibinfo {volume} {90}},\ \bibinfo
  {pages} {032132} (\bibinfo {year} {2014})},\ \Eprint
  {https://arxiv.org/abs/arXiv:1402.6362} {arXiv:1402.6362} \BibitemShut
  {NoStop}%
\bibitem [{\citenamefont {Gamayun}(2014)}]{gamayun_quantum_boltzmann_14}%
  \BibitemOpen
  \bibfield  {author} {\bibinfo {author} {\bibfnamefont {O.}~\bibnamefont
  {Gamayun}},\ }\bibfield  {title} {\bibinfo {title} {{Quantum Boltzmann
  equation for a mobile impurity in a degenerate Tonks-Girardeau gas}},\ }\href
  {https://doi.org/10.1103/PhysRevA.89.063627} {\bibfield  {journal} {\bibinfo
  {journal} {Phys. Rev. E}\ }\textbf {\bibinfo {volume} {89}},\ \bibinfo
  {pages} {063627} (\bibinfo {year} {2014})},\ \Eprint
  {https://arxiv.org/abs/arXiv:1402.7064} {arXiv:1402.7064} \BibitemShut
  {NoStop}%
\bibitem [{\citenamefont {McGuire}(1965)}]{mcguire1965interacting}%
  \BibitemOpen
  \bibfield  {author} {\bibinfo {author} {\bibfnamefont {J.~B.}\ \bibnamefont
  {McGuire}},\ }\bibfield  {title} {\bibinfo {title} {Interacting fermions in
  one dimension. i. repulsive potential},\ }\href
  {https://doi.org/10.1063/1.1704291} {\bibfield  {journal} {\bibinfo
  {journal} {J. Math. Phys.}\ }\textbf {\bibinfo {volume} {6}},\ \bibinfo
  {pages} {432} (\bibinfo {year} {1965})}\BibitemShut {NoStop}%
\bibitem [{\citenamefont {McGuire}(1966)}]{mcguire1966interacting}%
  \BibitemOpen
  \bibfield  {author} {\bibinfo {author} {\bibfnamefont {J.~B.}\ \bibnamefont
  {McGuire}},\ }\bibfield  {title} {\bibinfo {title} {Interacting fermions in
  one dimension. ii. attractive potential},\ }\href
  {https://doi.org/10.1063/1.1704798} {\bibfield  {journal} {\bibinfo
  {journal} {J. Math. Phys.}\ }\textbf {\bibinfo {volume} {7}},\ \bibinfo
  {pages} {123} (\bibinfo {year} {1966})}\BibitemShut {NoStop}%
\bibitem [{\citenamefont {Yang}(1967)}]{yang_fermions_spinful_67}%
  \BibitemOpen
  \bibfield  {author} {\bibinfo {author} {\bibfnamefont {C.~N.}\ \bibnamefont
  {Yang}},\ }\bibfield  {title} {\bibinfo {title} {{Some Exact Results for the
  Many-Body Problem in one Dimension with Repulsive Delta-Function
  Interaction}},\ }\href {https://doi.org/10.1103/PhysRevLett.19.1312}
  {\bibfield  {journal} {\bibinfo  {journal} {Phys. Rev. Lett.}\ }\textbf
  {\bibinfo {volume} {19}},\ \bibinfo {pages} {1312} (\bibinfo {year}
  {1967})}\BibitemShut {NoStop}%
\bibitem [{\citenamefont {Gamayun}\ \emph {et~al.}(2016)\citenamefont
  {Gamayun}, \citenamefont {Pronko},\ and\ \citenamefont
  {Zvonarev}}]{Gamayun_correlation_2016}%
  \BibitemOpen
  \bibfield  {author} {\bibinfo {author} {\bibfnamefont {O.}~\bibnamefont
  {Gamayun}}, \bibinfo {author} {\bibfnamefont {A.~G.}\ \bibnamefont
  {Pronko}},\ and\ \bibinfo {author} {\bibfnamefont {M.~B.}\ \bibnamefont
  {Zvonarev}},\ }\bibfield  {title} {\bibinfo {title} {Time and
  temperature-dependent correlation function of an impurity in one-dimensional
  fermi and tonks{\textendash}girardeau gases as a fredholm determinant},\
  }\href {https://doi.org/10.1088/1367-2630/18/4/045005} {\bibfield  {journal}
  {\bibinfo  {journal} {New J. Phys.}\ }\textbf {\bibinfo {volume} {18}},\
  \bibinfo {pages} {045005} (\bibinfo {year} {2016})},\ \Eprint
  {https://arxiv.org/abs/arXiv:1511.05922} {arXiv:1511.05922} \BibitemShut
  {NoStop}%
\bibitem [{\citenamefont {Gamayun}\ \emph {et~al.}(2018)\citenamefont
  {Gamayun}, \citenamefont {Lychkovskiy}, \citenamefont {Burovski},
  \citenamefont {Malcomson}, \citenamefont {Cheianov},\ and\ \citenamefont
  {Zvonarev}}]{gamayun_impact_18}%
  \BibitemOpen
  \bibfield  {author} {\bibinfo {author} {\bibfnamefont {O.}~\bibnamefont
  {Gamayun}}, \bibinfo {author} {\bibfnamefont {O.}~\bibnamefont
  {Lychkovskiy}}, \bibinfo {author} {\bibfnamefont {E.}~\bibnamefont
  {Burovski}}, \bibinfo {author} {\bibfnamefont {M.}~\bibnamefont {Malcomson}},
  \bibinfo {author} {\bibfnamefont {V.~V.}\ \bibnamefont {Cheianov}},\ and\
  \bibinfo {author} {\bibfnamefont {M.~B.}\ \bibnamefont {Zvonarev}},\
  }\bibfield  {title} {\bibinfo {title} {Impact of the injection protocol on an
  impurity's stationary state},\ }\href
  {https://doi.org/10.1103/PhysRevLett.120.220605} {\bibfield  {journal}
  {\bibinfo  {journal} {Phys. Rev. Lett.}\ }\textbf {\bibinfo {volume} {120}},\
  \bibinfo {pages} {220605} (\bibinfo {year} {2018})},\ \Eprint
  {https://arxiv.org/abs/arXiv:1708.07665} {arXiv:1708.07665} \BibitemShut
  {NoStop}%
\bibitem [{\citenamefont {Gamayun}\ \emph {et~al.}(2020)\citenamefont
  {Gamayun}, \citenamefont {Lychkovskiy},\ and\ \citenamefont
  {Zvonarev}}]{gamayun2019zero}%
  \BibitemOpen
  \bibfield  {author} {\bibinfo {author} {\bibfnamefont {O.}~\bibnamefont
  {Gamayun}}, \bibinfo {author} {\bibfnamefont {O.}~\bibnamefont
  {Lychkovskiy}},\ and\ \bibinfo {author} {\bibfnamefont {M.~B.}\ \bibnamefont
  {Zvonarev}},\ }\bibfield  {title} {\bibinfo {title} {{Zero temperature
  momentum distribution of an impurity in a polaron state of one-dimensional
  Fermi and Tonks-Girardeau gases}},\ }\href
  {https://doi.org/10.21468/SciPostPhys.8.4.053} {\bibfield  {journal}
  {\bibinfo  {journal} {SciPost Phys.}\ }\textbf {\bibinfo {volume} {8}},\
  \bibinfo {pages} {53} (\bibinfo {year} {2020})},\ \Eprint
  {https://arxiv.org/abs/arXiv:1909.07358} {arXiv:1909.07358} \BibitemShut
  {NoStop}%
\bibitem [{\citenamefont {Imambekov}\ \emph {et~al.}(2012)\citenamefont
  {Imambekov}, \citenamefont {Schmidt},\ and\ \citenamefont
  {Glazman}}]{imambekov_review_12}%
  \BibitemOpen
  \bibfield  {author} {\bibinfo {author} {\bibfnamefont {A.}~\bibnamefont
  {Imambekov}}, \bibinfo {author} {\bibfnamefont {T.~L.}\ \bibnamefont
  {Schmidt}},\ and\ \bibinfo {author} {\bibfnamefont {L.~I.}\ \bibnamefont
  {Glazman}},\ }\bibfield  {title} {\bibinfo {title} {{One-dimensional quantum
  liquids: Beyond the Luttinger liquid paradigm}},\ }\href
  {https://doi.org/10.1103/RevModPhys.84.1253} {\bibfield  {journal} {\bibinfo
  {journal} {Rev. Mod. Phys.}\ }\textbf {\bibinfo {volume} {84}},\ \bibinfo
  {pages} {1253} (\bibinfo {year} {2012})},\ \Eprint
  {https://arxiv.org/abs/arXiv:1110.1374} {arXiv:1110.1374} \BibitemShut
  {NoStop}%
\bibitem [{\citenamefont {Mathy}\ \emph {et~al.}(2012)\citenamefont {Mathy},
  \citenamefont {Zvonarev},\ and\ \citenamefont {Demler}}]{mathy2012quantum}%
  \BibitemOpen
  \bibfield  {author} {\bibinfo {author} {\bibfnamefont {C.~J.~M.}\
  \bibnamefont {Mathy}}, \bibinfo {author} {\bibfnamefont {M.~B.}\ \bibnamefont
  {Zvonarev}},\ and\ \bibinfo {author} {\bibfnamefont {E.}~\bibnamefont
  {Demler}},\ }\bibfield  {title} {\bibinfo {title} {Quantum flutter of
  supersonic particles in one-dimensional quantum liquids},\ }\href
  {https://doi.org/10.1038/nphys2455} {\bibfield  {journal} {\bibinfo
  {journal} {Nat. Phys.}\ }\textbf {\bibinfo {volume} {8}},\ \bibinfo {pages}
  {881} (\bibinfo {year} {2012})},\ \Eprint
  {https://arxiv.org/abs/arXiv:1203.4819} {arXiv:1203.4819} \BibitemShut
  {NoStop}%
\bibitem [{\citenamefont {Hackl}\ \emph {et~al.}(2020)\citenamefont {Hackl},
  \citenamefont {Guaita}, \citenamefont {Shi}, \citenamefont {Haegeman},
  \citenamefont {Demler},\ and\ \citenamefont {Cirac}}]{hackl2020geometry}%
  \BibitemOpen
  \bibfield  {author} {\bibinfo {author} {\bibfnamefont {L.}~\bibnamefont
  {Hackl}}, \bibinfo {author} {\bibfnamefont {T.}~\bibnamefont {Guaita}},
  \bibinfo {author} {\bibfnamefont {T.}~\bibnamefont {Shi}}, \bibinfo {author}
  {\bibfnamefont {J.}~\bibnamefont {Haegeman}}, \bibinfo {author}
  {\bibfnamefont {E.}~\bibnamefont {Demler}},\ and\ \bibinfo {author}
  {\bibfnamefont {I.}~\bibnamefont {Cirac}},\ }\bibfield  {title} {\bibinfo
  {title} {Geometry of variational methods: dynamics of closed quantum
  systems},\ }\href@noop {} {\  (\bibinfo {year} {2020})},\ \Eprint
  {https://arxiv.org/abs/arXiv:2004.01015} {arXiv:2004.01015} \BibitemShut
  {NoStop}%
\bibitem [{\citenamefont {Shi}\ \emph {et~al.}(2019)\citenamefont {Shi},
  \citenamefont {Demler},\ and\ \citenamefont {Cirac}}]{shi2019variational}%
  \BibitemOpen
  \bibfield  {author} {\bibinfo {author} {\bibfnamefont {T.}~\bibnamefont
  {Shi}}, \bibinfo {author} {\bibfnamefont {E.}~\bibnamefont {Demler}},\ and\
  \bibinfo {author} {\bibfnamefont {J.~I.}\ \bibnamefont {Cirac}},\ }\bibfield
  {title} {\bibinfo {title} {A variational approach for many-body systems at
  finite temperature},\ }\href@noop {} {\bibfield  {journal} {\bibinfo
  {journal} {arXiv preprint arXiv:1912.11907}\ } (\bibinfo {year}
  {2019})}\BibitemShut {NoStop}%
\bibitem [{\citenamefont {Lee}\ \emph {et~al.}(1953)\citenamefont {Lee},
  \citenamefont {Low},\ and\ \citenamefont {Pines}}]{lee1953motion}%
  \BibitemOpen
  \bibfield  {author} {\bibinfo {author} {\bibfnamefont {T.~D.}\ \bibnamefont
  {Lee}}, \bibinfo {author} {\bibfnamefont {F.~E.}\ \bibnamefont {Low}},\ and\
  \bibinfo {author} {\bibfnamefont {D.}~\bibnamefont {Pines}},\ }\bibfield
  {title} {\bibinfo {title} {The motion of slow electrons in a polar crystal},\
  }\href {https://doi.org/10.1103/PhysRev.90.297} {\bibfield  {journal}
  {\bibinfo  {journal} {Phys. Rev.}\ }\textbf {\bibinfo {volume} {90}},\
  \bibinfo {pages} {297} (\bibinfo {year} {1953})}\BibitemShut {NoStop}%
\bibitem [{\citenamefont {Kamenev}(2011)}]{kamenev2011field}%
  \BibitemOpen
  \bibfield  {author} {\bibinfo {author} {\bibfnamefont {A.}~\bibnamefont
  {Kamenev}},\ }\bibfield  {title} {\bibinfo {title} {Field theory of
  non-equilibrium systems},\ }\href@noop {} {\bibfield  {journal} {\bibinfo
  {journal} {Field Theory of Non-Equilibrium Systems, by Alex Kamenev,
  Cambridge, UK: Cambridge University Press, 2011}\ } (\bibinfo {year}
  {2011})}\BibitemShut {NoStop}%
\bibitem [{\citenamefont {Guaita}\ \emph {et~al.}(2019)\citenamefont {Guaita},
  \citenamefont {Hackl}, \citenamefont {Shi}, \citenamefont {Hubig},
  \citenamefont {Demler},\ and\ \citenamefont {Cirac}}]{guaita2019gaussian}%
  \BibitemOpen
  \bibfield  {author} {\bibinfo {author} {\bibfnamefont {T.}~\bibnamefont
  {Guaita}}, \bibinfo {author} {\bibfnamefont {L.}~\bibnamefont {Hackl}},
  \bibinfo {author} {\bibfnamefont {T.}~\bibnamefont {Shi}}, \bibinfo {author}
  {\bibfnamefont {C.}~\bibnamefont {Hubig}}, \bibinfo {author} {\bibfnamefont
  {E.}~\bibnamefont {Demler}},\ and\ \bibinfo {author} {\bibfnamefont {J.~I.}\
  \bibnamefont {Cirac}},\ }\bibfield  {title} {\bibinfo {title} {Gaussian
  time-dependent variational principle for the bose-hubbard model},\ }\href
  {https://doi.org/10.1103/PhysRevB.100.094529} {\bibfield  {journal} {\bibinfo
   {journal} {Phys. Rev. B}\ }\textbf {\bibinfo {volume} {100}},\ \bibinfo
  {pages} {094529} (\bibinfo {year} {2019})},\ \Eprint
  {https://arxiv.org/abs/arXiv:1907.04837} {arXiv:1907.04837} \BibitemShut
  {NoStop}%
\bibitem [{\citenamefont {Knap}\ \emph {et~al.}(2012)\citenamefont {Knap},
  \citenamefont {Shashi}, \citenamefont {Nishida}, \citenamefont {Imambekov},
  \citenamefont {Abanin},\ and\ \citenamefont {Demler}}]{knap2012time}%
  \BibitemOpen
  \bibfield  {author} {\bibinfo {author} {\bibfnamefont {M.}~\bibnamefont
  {Knap}}, \bibinfo {author} {\bibfnamefont {A.}~\bibnamefont {Shashi}},
  \bibinfo {author} {\bibfnamefont {Y.}~\bibnamefont {Nishida}}, \bibinfo
  {author} {\bibfnamefont {A.}~\bibnamefont {Imambekov}}, \bibinfo {author}
  {\bibfnamefont {D.~A.}\ \bibnamefont {Abanin}},\ and\ \bibinfo {author}
  {\bibfnamefont {E.}~\bibnamefont {Demler}},\ }\bibfield  {title} {\bibinfo
  {title} {Time-dependent impurity in ultracold fermions: Orthogonality
  catastrophe and beyond},\ }\href {https://doi.org/10.1103/PhysRevX.2.041020}
  {\bibfield  {journal} {\bibinfo  {journal} {Phys. Rev. X}\ }\textbf {\bibinfo
  {volume} {2}},\ \bibinfo {pages} {041020} (\bibinfo {year}
  {2012})}\BibitemShut {NoStop}%
\bibitem [{\citenamefont {Ness}\ \emph {et~al.}(2020)\citenamefont {Ness},
  \citenamefont {Shkedrov}, \citenamefont {Florshaim}, \citenamefont {Diessel},
  \citenamefont {von Milczewski}, \citenamefont {Schmidt},\ and\ \citenamefont
  {Sagi}}]{ness2020observation}%
  \BibitemOpen
  \bibfield  {author} {\bibinfo {author} {\bibfnamefont {G.}~\bibnamefont
  {Ness}}, \bibinfo {author} {\bibfnamefont {C.}~\bibnamefont {Shkedrov}},
  \bibinfo {author} {\bibfnamefont {Y.}~\bibnamefont {Florshaim}}, \bibinfo
  {author} {\bibfnamefont {O.~K.}\ \bibnamefont {Diessel}}, \bibinfo {author}
  {\bibfnamefont {J.}~\bibnamefont {von Milczewski}}, \bibinfo {author}
  {\bibfnamefont {R.}~\bibnamefont {Schmidt}},\ and\ \bibinfo {author}
  {\bibfnamefont {Y.}~\bibnamefont {Sagi}},\ }\bibfield  {title} {\bibinfo
  {title} {Observation of a smooth polaron-molecule transition in a degenerate
  fermi gas},\ }\href@noop {} {\bibfield  {journal} {\bibinfo  {journal} {arXiv
  preprint arXiv:2001.10450}\ } (\bibinfo {year} {2020})}\BibitemShut {NoStop}%
\bibitem [{\citenamefont {Stenger}\ \emph {et~al.}(1999)\citenamefont
  {Stenger}, \citenamefont {Inouye}, \citenamefont {Chikkatur}, \citenamefont
  {Stamper-Kurn}, \citenamefont {Pritchard},\ and\ \citenamefont
  {Ketterle}}]{stenger1999bragg}%
  \BibitemOpen
  \bibfield  {author} {\bibinfo {author} {\bibfnamefont {J.}~\bibnamefont
  {Stenger}}, \bibinfo {author} {\bibfnamefont {S.}~\bibnamefont {Inouye}},
  \bibinfo {author} {\bibfnamefont {A.~P.}\ \bibnamefont {Chikkatur}}, \bibinfo
  {author} {\bibfnamefont {D.~M.}\ \bibnamefont {Stamper-Kurn}}, \bibinfo
  {author} {\bibfnamefont {D.~E.}\ \bibnamefont {Pritchard}},\ and\ \bibinfo
  {author} {\bibfnamefont {W.}~\bibnamefont {Ketterle}},\ }\bibfield  {title}
  {\bibinfo {title} {Bragg spectroscopy of a bose-einstein condensate},\ }\href
  {https://doi.org/10.1103/PhysRevLett.82.4569} {\bibfield  {journal} {\bibinfo
   {journal} {Phys. Rev. Lett.}\ }\textbf {\bibinfo {volume} {82}},\ \bibinfo
  {pages} {4569} (\bibinfo {year} {1999})}\BibitemShut {NoStop}%
\bibitem [{\citenamefont {Stamper-Kurn}\ \emph {et~al.}(1999)\citenamefont
  {Stamper-Kurn}, \citenamefont {Chikkatur}, \citenamefont {G{\"o}rlitz},
  \citenamefont {Inouye}, \citenamefont {Gupta}, \citenamefont {Pritchard},\
  and\ \citenamefont {Ketterle}}]{stamper1999excitation}%
  \BibitemOpen
  \bibfield  {author} {\bibinfo {author} {\bibfnamefont {D.~M.}\ \bibnamefont
  {Stamper-Kurn}}, \bibinfo {author} {\bibfnamefont {A.~P.}\ \bibnamefont
  {Chikkatur}}, \bibinfo {author} {\bibfnamefont {A.}~\bibnamefont
  {G{\"o}rlitz}}, \bibinfo {author} {\bibfnamefont {S.}~\bibnamefont {Inouye}},
  \bibinfo {author} {\bibfnamefont {S.}~\bibnamefont {Gupta}}, \bibinfo
  {author} {\bibfnamefont {D.~E.}\ \bibnamefont {Pritchard}},\ and\ \bibinfo
  {author} {\bibfnamefont {W.}~\bibnamefont {Ketterle}},\ }\bibfield  {title}
  {\bibinfo {title} {Excitation of phonons in a bose-einstein condensate by
  light scattering},\ }\href {https://doi.org/10.1103/PhysRevLett.83.2876}
  {\bibfield  {journal} {\bibinfo  {journal} {Phys. Rev. Lett.}\ }\textbf
  {\bibinfo {volume} {83}},\ \bibinfo {pages} {2876} (\bibinfo {year}
  {1999})}\BibitemShut {NoStop}%
\bibitem [{\citenamefont {Ozeri}\ \emph {et~al.}(2005)\citenamefont {Ozeri},
  \citenamefont {Katz}, \citenamefont {Steinhauer},\ and\ \citenamefont
  {Davidson}}]{ozeri2005colloquium}%
  \BibitemOpen
  \bibfield  {author} {\bibinfo {author} {\bibfnamefont {R.}~\bibnamefont
  {Ozeri}}, \bibinfo {author} {\bibfnamefont {N.}~\bibnamefont {Katz}},
  \bibinfo {author} {\bibfnamefont {J.}~\bibnamefont {Steinhauer}},\ and\
  \bibinfo {author} {\bibfnamefont {N.}~\bibnamefont {Davidson}},\ }\bibfield
  {title} {\bibinfo {title} {Colloquium: Bulk bogoliubov excitations in a
  bose-einstein condensate},\ }\href
  {https://doi.org/10.1103/RevModPhys.77.187} {\bibfield  {journal} {\bibinfo
  {journal} {Rev. Mod. Phys.}\ }\textbf {\bibinfo {volume} {77}},\ \bibinfo
  {pages} {187} (\bibinfo {year} {2005})}\BibitemShut {NoStop}%
\bibitem [{\citenamefont {Cohen-Tannoudji}\ \emph {et~al.}(1998)\citenamefont
  {Cohen-Tannoudji}, \citenamefont {Dupont-Roc},\ and\ \citenamefont
  {Grynberg}}]{cohen1998atom}%
  \BibitemOpen
  \bibfield  {author} {\bibinfo {author} {\bibfnamefont {C.}~\bibnamefont
  {Cohen-Tannoudji}}, \bibinfo {author} {\bibfnamefont {J.}~\bibnamefont
  {Dupont-Roc}},\ and\ \bibinfo {author} {\bibfnamefont {G.}~\bibnamefont
  {Grynberg}},\ }\href@noop {} {\emph {\bibinfo {title} {Atom-photon
  interactions: basic processes and applications}}}\ (\bibinfo {year}
  {1998})\BibitemShut {NoStop}%
\end{thebibliography}%

\appendix

\section{Benchmarking the ground-state correlations with the Bethe ansatz results}
\label{Appendix_BA_static}

\begin{figure}[b!]
\centering
\includegraphics[width=0.9\linewidth]{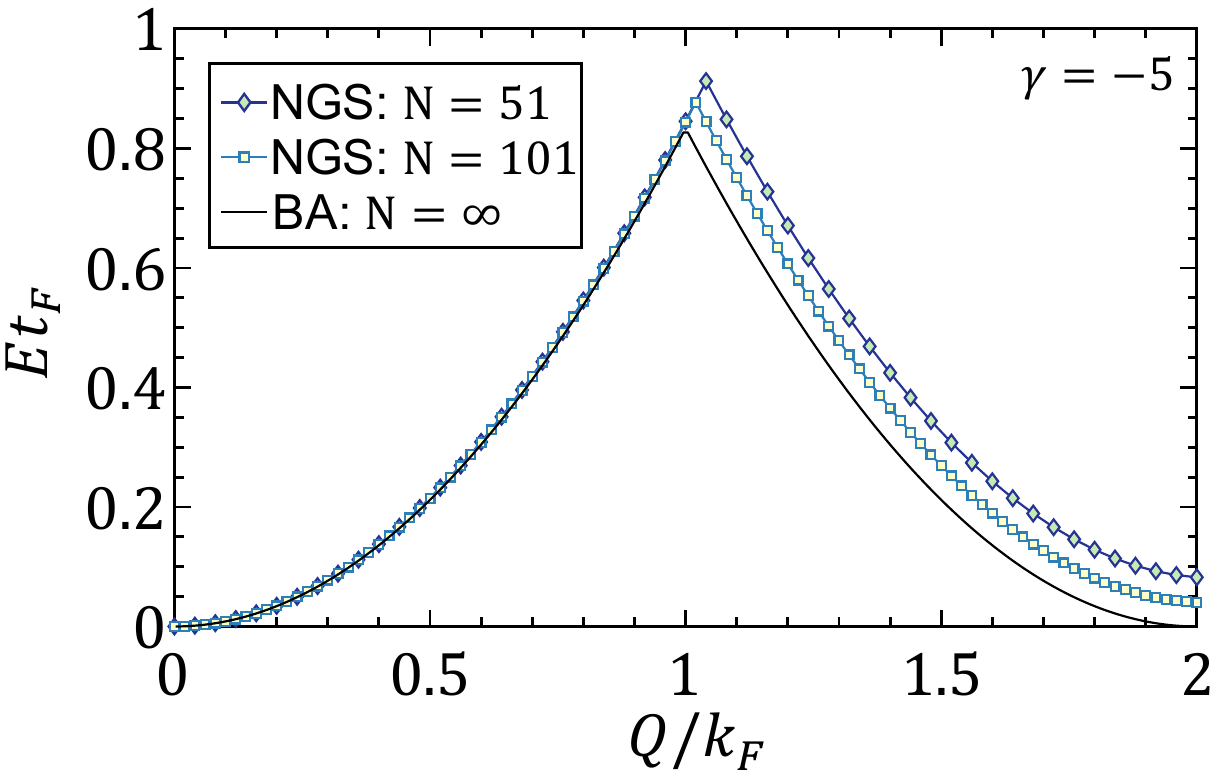} 
\caption{Polaron energy-momentum relation for the case of attractive coupling $\gamma < 0$ (here we fix $\Lambda = 10 k_F$). We find that our variational calculation nicely reproduces the BA result, adopted from Refs.~\cite{mcguire1966interacting,gamayun2019zero}.}
\label{fig::BA_cmp_EM_attr}
\end{figure}

In the main text, we showed that for the case of repulsive coupling $\gamma > 0$, the non-Gaussian variational approach reliably reproduces exact ground-state energies, available from the BA for $M=m$. Figure~\ref{fig::BA_cmp_EM_attr} demonstrates that our method also agrees with the BA for the case of attractive coupling $\gamma < 0$. Remarkably, even the kink-like feature at $Q = k_F$ is reproduced. Below we further compute various many-body correlation functions via the variational approach and compare them to the BA results.

\begin{figure}[t!]
\centering
\includegraphics[width=1\linewidth]{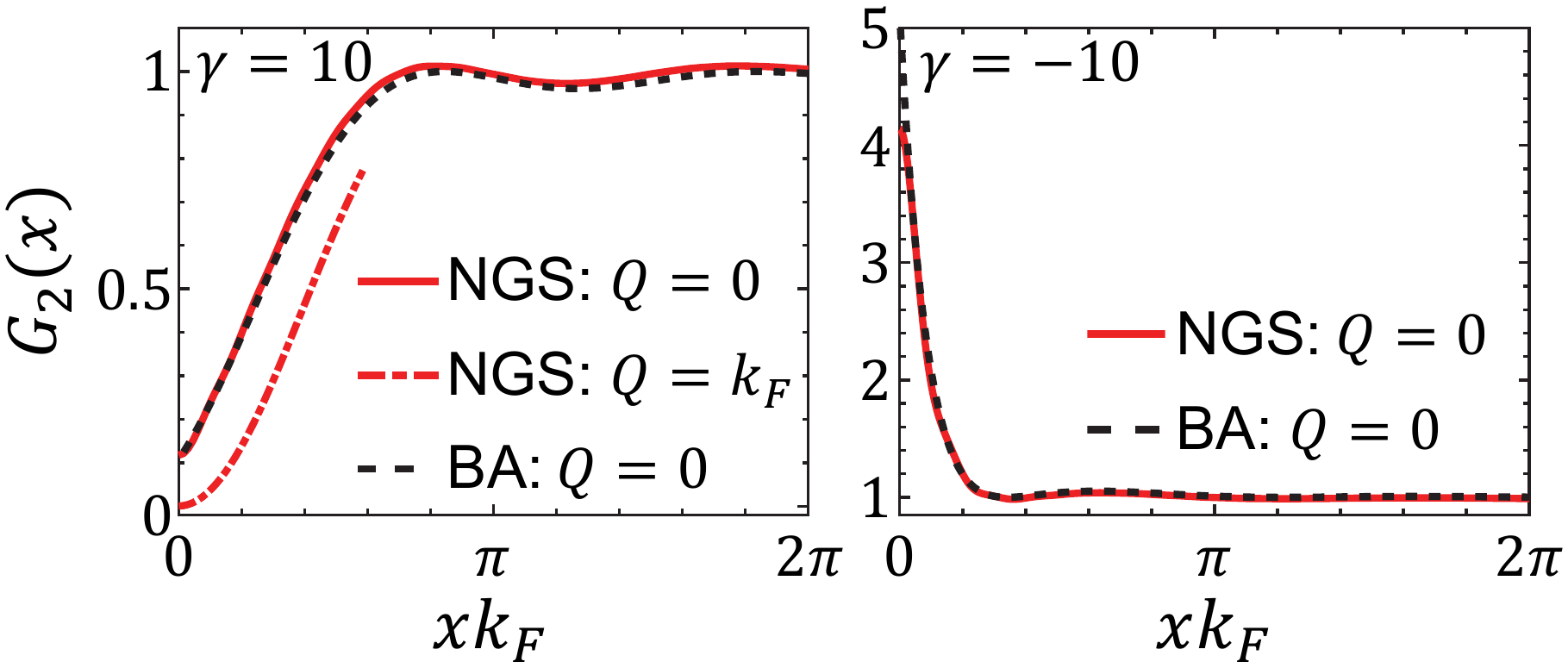} 
\caption{Comparison of the two-point static correlation function $G_2(x)$: solid (red) curves correspond to the non-Gaussian variational approach; dashed (black) lines represent the BA analytical results adopted from Refs.~\cite{mcguire1965interacting,mcguire1966interacting}. We consider both repulsion (left panel) and attraction (right panel). Parameters used: $\Lambda = 15k_F$ and $N = 51.$}
\label{fig::BA_cmp_correlations_G2}
\end{figure}

We first compute the two-point correlation function, $G_2(x)$, defined in the main text. This correlator describes the probability density to find a bath particle separated from the impurity by the distance $x$. Within the Gaussian variational approach, it is obtained as
\begin{eqnarray}
    G_2(x) =  \frac{L}{N}  \langle\hat{c}^\dagger_{x}\hat{c}_{x}\rangle_{\rm GS}.\label{eqn::G2_VM}
\end{eqnarray}
Figure~\ref{fig::BA_cmp_correlations_G2} shows the comparison of our calculation~\eqref{eqn::G2_VM} to the BA result adopted from Refs.~\cite{mcguire1965interacting,mcguire1966interacting}, and we observe an excellent agreement for both repulsive and attractive couplings.

\begin{figure}[b!]
\centering
\includegraphics[width=1\linewidth]{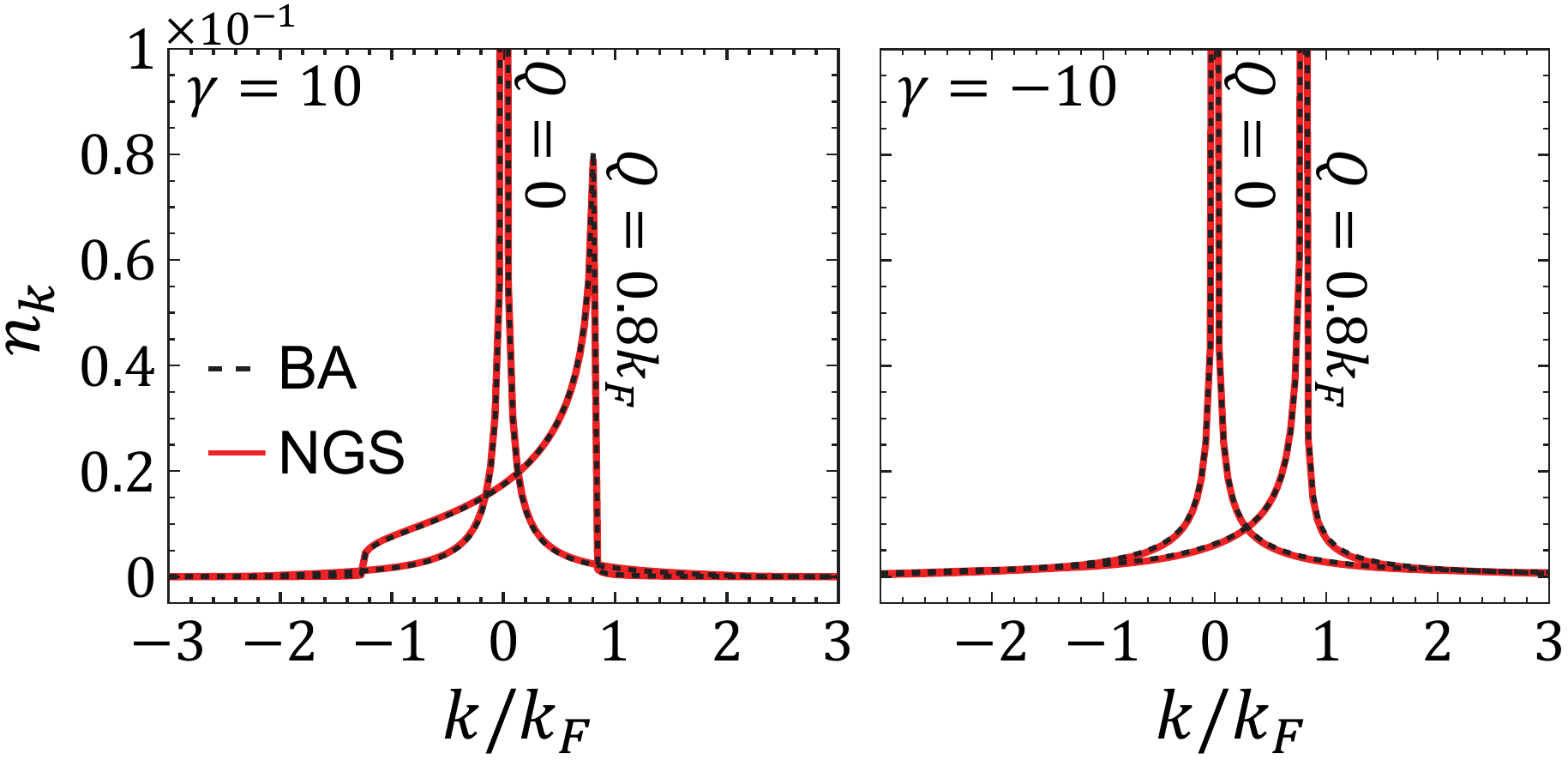} 
\caption{Comparison of the impurity momentum distribution function $n_k(Q)$: solid (red) curves correspond to simulations of Eq.~\eqref{eqn:n_k}; the BA results (dashed black lines) are taken from Ref.~\cite{gamayun2019zero}. Parameters are the same as in Fig.~\ref{fig::BA_cmp_correlations_G2}.}
\label{fig::BA_cmp_correlations_nk}
\end{figure}

\begin{figure}[b!]
\centering
\includegraphics[width=1\linewidth]{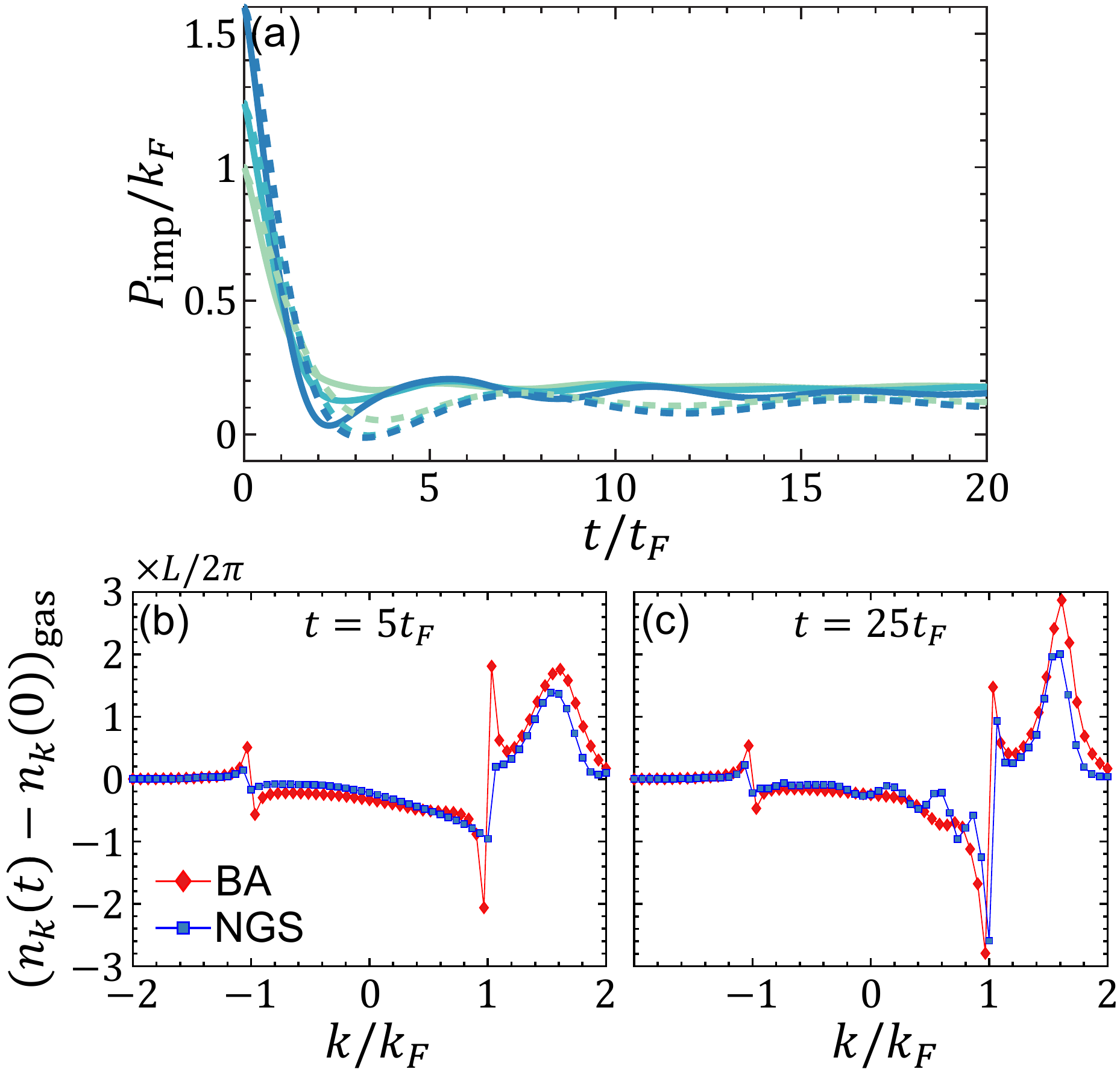} 
\caption{Comparison of out-of-equilibrium correlations. (a) Evolution of the impurity momentum for three different initial conditions. Solid curves represent our variational simulations and dashed curves show the BA results adopted from Ref.~\cite{mathy2012quantum}. Note (i) damped quantum-flutter oscillations for intermediate times and (ii) saturation of the impurity momentum at $t\to \infty$. The non-Gaussian approach well captures the initial dynamics and saturation, but the discrepancy at intermediate times is clear. Parameters used: $\gamma = 10$, $N = 51$, $\Lambda = 10 k_F$. (b) and (c) Momentum distribution function of the host-gas particles at $t = 5 t_F$ (b) and $t = 25 t_F$ (c). These graphs demonstrate that dynamical NGS correlations are in good agreement with those of the BA~\cite{mathy2012quantum}. Parameters used: $\gamma = 5$, $N = 31$, $\Lambda = 10 k_F$, $P_{\rm imp}(0) \approx 1.35 k_F$.}
\label{fig::BA_cmp_dynamics}
\end{figure}

 Another interesting observable to calculate is the momentum distribution function of the impurity in the laboratory frame:
\begin{align}
    n_k(Q) & = \Bra{\psi_{\rm lab}}\hat{d}_k^\dagger \hat{d}_k\Ket{\psi_{\rm lab}} = \Bra{\psi_{\rm LLP}} \hat{d}_{k +\hat{P}_f}^\dagger \hat{d}_{k +\hat{P}_f}\Ket{\psi_{\rm LLP}} \notag \\
    &\qquad = \int\limits_{0}^{L}\frac{dx}{L} \Big\langle\exp{\Big(ix(k+\hat{P}_f - Q\Big) }\Big\rangle_{\rm GS}.
\end{align}
Using the formalism of Ref.~\cite{shi2018variational}, the latter expectation value can be computed analytically:
\begin{align}
    n_k = \int\limits_{0}^{L}\frac{dx}{L} {\rm e}^{i(k-Q)x} \det \left[ \left({\rm e}^{i\hat{K}x} - \hat{1} \right)  \Gamma_{Q}  + \hat{1} \right], \label{eqn:n_k}
\end{align}
where $\hat{K}_{k,k'} = k\delta_{k,k'}$. This integral we evaluate numerically once the covariance matrix $\Gamma_{Q}$ has been computed using the imaginary-time evolution. In Fig.~\ref{fig::BA_cmp_correlations_nk}, we compare our approach with the BA calculation of Ref.~\cite{gamayun2019zero}: The agreement between the methods is quite striking. We, therefore, conclude that our non-Gaussian approach not only predicts correctly ground-state energies but also captures properties of wave-function correlations.

\section{Benchmarking far-from-equilibrium dynamics}
\label{Appendix_dyn}

We now aim to test our real-time approach, encoded in Eq.~\eqref{eqn::real_dyn_Gamma}, and apply our formalism to the so-called quantum flutter outlined below. Importantly, recent exact Bethe ansatz~\cite{mathy2012quantum} calculations and simulations with matrix product states~\cite{knap2014quantum} provide a necessary ground to benchmark our variational method. 

We choose the following initial state for the real-time dynamics:
\begin{equation}
    \Ket{\Psi^{\rm lab}_{ Q}} = \Ket{\rm FS}\otimes\Ket{Q}_{\rm imp}.
\end{equation}
Note that the total momentum of the system is $Q$ and, hence, in the co-moving frame, we can stick to a single momentum sector.

In Fig.~\ref{fig::BA_cmp_dynamics}\,(a), we plot the evolution of the impurity momentum $P_{\rm imp}(t)$ and compare our results with the BA calculations of Ref.~\cite{mathy2012quantum}. The dynamics exhibits three stages: (i) initial rapid decay during which the impurity redistributes its momentum to the host-gas particles; (ii) intermediate-time oscillations called quantum flutter; and (iii) saturation to a steady-state with a non-zero impurity momentum. We see that our method captures the three stages correctly, though, there is a clear discrepancy compared to the exact result at intermediate times. This mismatch indicates that our variational wave function is too restrictive to reproduce full transient dynamics quantitatively. At the same time, the essential physics is well reproduced qualitatively. Furthermore, Fig.~\ref{fig::BA_cmp_dynamics}\,(b) and (c) shows that the evolution of the fermionic momentum distribution function computed with the NGS approach matches well the BA results.

\begin{figure}[t!]
\centering
\includegraphics[width=1\linewidth]{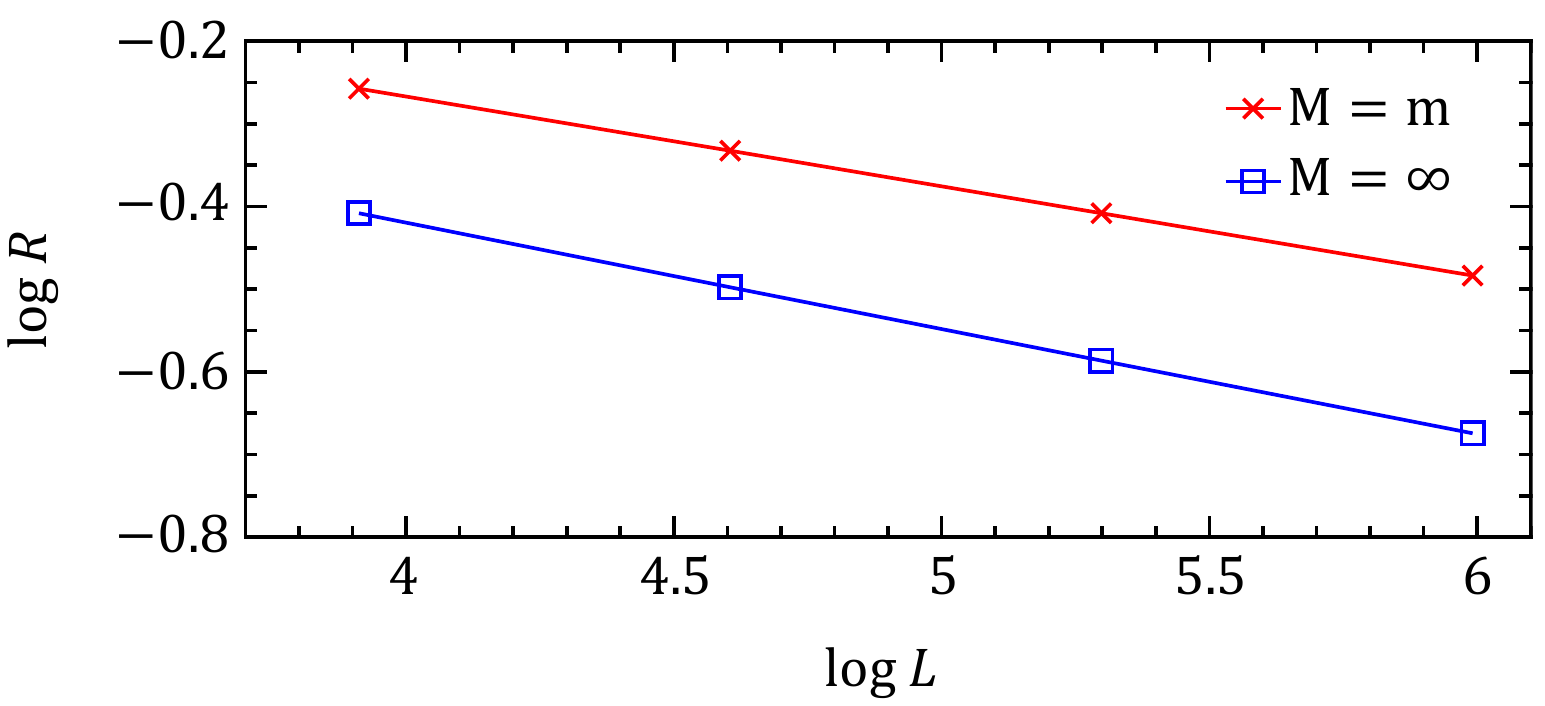} 
\caption{Quasiparticle residue as a function of the system size, $L$. The density of fermions is kept constant. Parameters used: $\gamma = 5$ and $\Lambda = 5k_F$. }
\label{fig::Residue}
\end{figure}

\begin{figure*}[t!]
\centering
\includegraphics[width=0.9\linewidth]{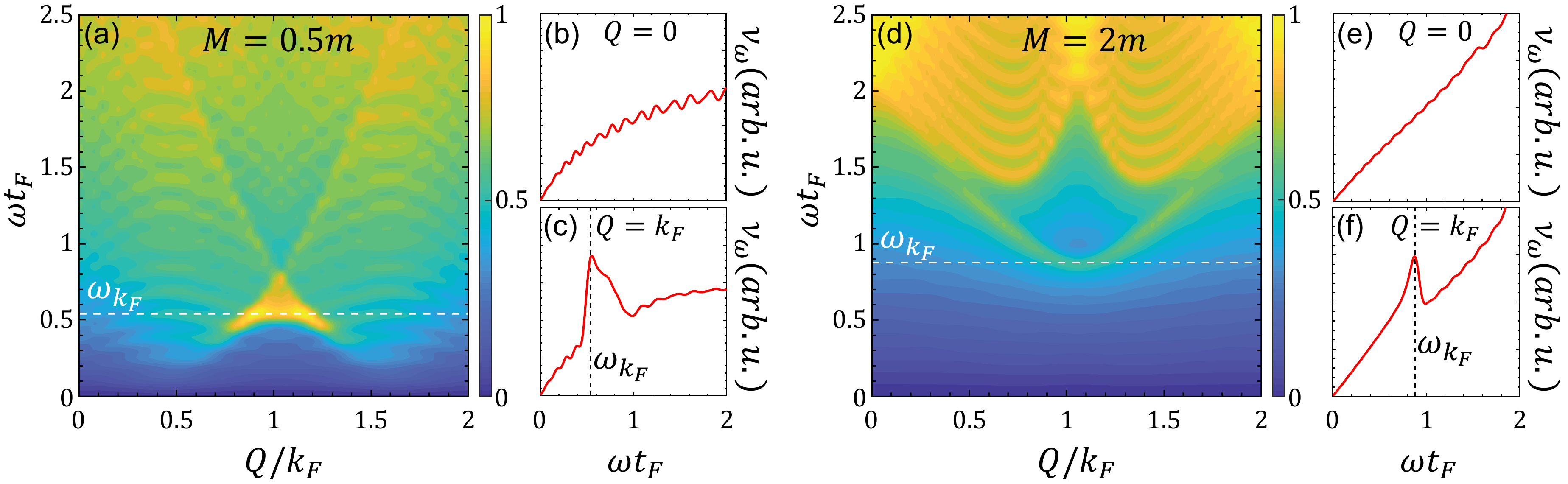} 
\caption{The DOS of collective excitations as a function of frequency $\omega$ and total momentum $Q$ for $M = 0.5 m$ (panels a, b, and c) and for $M = 2 m$ (panels d, e, and f). For $M = 0.5m$, we observe a rather dispersive peak; at $Q = k_F$, the DOS is notably broad (c). This is because the magnon branch approaches plasmon for small impurity masses so that their dispersions are parallel to each other over a wide range of momenta near $k_F$. For $M = 2m$, the peak at $Q = k_F$ is sharp (f), but its spectral weight is relatively weak. Parameters are the same as in Fig.~\ref{fig::Main}.}
\label{fig::Figure_two_cuts}
\end{figure*}

\begin{figure}[t!]
\centering
\includegraphics[width=1\linewidth]{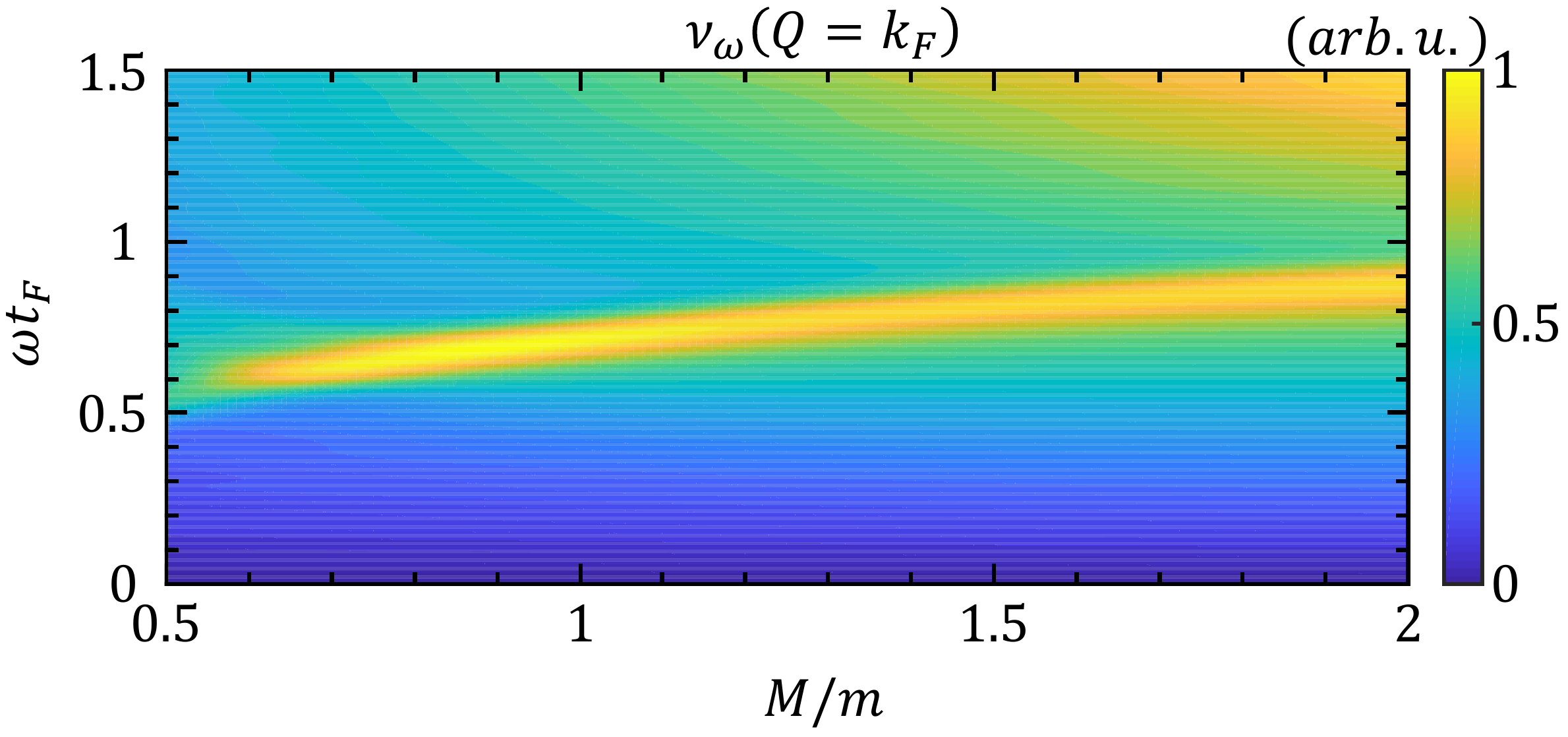} 
\caption{The DOS at $Q = k_F$ as a function of frequency $\omega$ and mass ratio $M/m$.  Parameters are the same as in Fig.~\ref{fig::Main}.}
\label{fig::Figure_full}
\end{figure}

\section{Quasiparticle residue}
\label{Appendix_Residue}

Within the Gaussian-states framework, the quasiparticle residue, defined as $R = \lvert \braket{\rm FS| \psi_{\rm GS}}\rvert^2$ for $Q = 0$, can be easily computed:
\begin{align}
    R = \det(\hat{1} + 2{\Gamma}_{\rm GS}{\Gamma}_{\rm FS} - ({\Gamma}_{\rm GS}+{\Gamma}_{\rm FS})),
\end{align}
where ${\Gamma}_{\rm FS}$ is the covariance matrix of the filled Fermi sea. The residue $R$ becomes suppressed as a power law of the system size $L$ -- see Fig.~\ref{fig::Residue}. For $M=\infty$, this result is known as the Anderson orthogonality catastrophe. Due to the non-vanishing recoil energy in Eq.~\eqref{eqn::LLP_H}, the suppression of the residue for the case of mobile impurity $M = m$ is slower compared to the case of infinitely heavy impurity.

\section{Collective modes for $M\neq m$}
\label{appendix:CM}

In this appendix, we provide more discussion of the properties of the sharp mode for $M \neq m$.  Figure~\ref{fig::Figure_two_cuts} shows the DOS of collective excitations for $M = 0.5 m$ (left panels) and $M = 2m$ (right panels). The case of heavy impurities qualitatively reminds the $M = m$ result in Fig.~\ref{fig::Main}~(b): We observe the development of a sharp peak at $Q = k_F$, but its relative spectral weight decreases with increasing the impurity mass -- see also Fig.~\ref{fig::Figure_full}, which shows the DOS at $k_F$ as a function of the mass ratio $M/m$. The case of light impurities is different. As the impurity mass decreases, the entire magnon branch approaches that of the plasmon so that their dispersions become nearly parallel to each other in a wide vicinity of $Q = k_F$. This, in turn, results in i) a rather dispersive peak in the DOS of collective excitations (Fig.~\ref{fig::Figure_two_cuts}~(a)) and ii) a substantial broadening at $k_F$ (Fig.~\ref{fig::Figure_two_cuts}~(c)). This latter broadening indicates that it is challenging to extract the quantum flutter frequency in dynamical simulations reliably.

\clearpage

\end{document}